\documentstyle[12pt,chicago]{article}

\newcommand{\cl}{{\it cl}}
\newcommand\pow[1]{{\cal P}(#1)}

\newcommand\rimp{\Rightarrow}
\newcommand{\liff}{\Leftrightarrow}
\newcommand\la{\langle}
\newcommand\ra{\rangle}
\newcommand\<{\langle}
\renewcommand\>{\rangle}

\newcommand\s{\mbox{\rm 's}\,}

\newcommand\cert{~{\it cert}~}

\newcommand\spk{\longmapsto} %

\newcommand\bdto{\spk} %

\newcommand{\commentout}[1]{}

\newcommand\be{\begin{enumerate}}
\newcommand\ee{\end{enumerate}}

\newcommand\bi{\begin{itemize}}
\newcommand\ei{\end{itemize}}

\newtheorem{THEOREM}{Theorem}[section]
\newenvironment{theorem}{\begin{THEOREM} \hspace{-.85em} {\bf :} }%
                        {\end{THEOREM}}
\newtheorem{LEMMA}[THEOREM]{Lemma}
\newenvironment{lemma}{\begin{LEMMA} \hspace{-.85em} {\bf :} }%
                      {\end{LEMMA}}
\newtheorem{COROLLARY}[THEOREM]{Corollary}
\newenvironment{corollary}{\begin{COROLLARY} \hspace{-.85em} {\bf :} }%
                          {\end{COROLLARY}}
\newtheorem{PROPOSITION}[THEOREM]{Proposition}
\newenvironment{proposition}{\begin{PROPOSITION} \hspace{-.85em} {\bf :}
}%
                            {\end{PROPOSITION}}
\newtheorem{DEFINITION}[THEOREM]{Definition}
\newenvironment{definition}{\begin{DEFINITION} \hspace{-.85em} {\bf :}
\rm}%
                            {\end{DEFINITION}}
\newtheorem{CLAIM}[THEOREM]{Claim}
                            {\end{CLAIM}}
\newtheorem{EXAMPLE}[THEOREM]{Example}
\newenvironment{example}{\begin{EXAMPLE} \hspace{-.85em} {\bf :} \rm}%
                            {\end{EXAMPLE}}
\newtheorem{REMARK}[THEOREM]{Remark}
\newenvironment{remark}{\begin{REMARK} \hspace{-.85em} {\bf :} \rm}%
                            {\end{REMARK}}
\newcommand{\thm}{\begin{theorem}}
\newcommand{\lem}{\begin{lemma}}
\newcommand{\pro}{\begin{proposition}}
\newcommand{\dfn}{\begin{definition}}
\newcommand{\rem}{\begin{remark}}
\newcommand{\xam}{\begin{example}}
\newcommand{\cor}{\begin{corollary}}
\newcommand{\prf}{\noindent{\bf Proof:} }
\newcommand{\ethm}{\end{theorem}}
\newcommand{\elem}{\end{lemma}}
\newcommand{\epro}{\end{proposition}}
\newcommand{\edfn}{\bbox\end{definition}}
\newcommand{\erem}{\bbox\end{remark}}
\newcommand{\exam}{\bbox\end{example}}

\newcommand{\ecor}{\end{corollary}}
\newcommand{\eprf}{\bbox\vspace{0.1in}}
\newcommand{\tn}{{\tt n}}
\newcommand{\tp}{{\tt p}}

\newcommand{\tk}{{\tt k}}
\newcommand{\tc}{{\tt c}}
\newcommand{\tm}{{\tt m}}
\newcommand{\ttt}{{\tt t}}
\newcommand{\tts}{{\tt t}}
\newcommand{\emptyint}{\emptyset}

\newcommand{\bbox}{\vrule height7pt width4pt depth1pt}

\newcommand{\intension}[2]{[\![ #1 ]\!]_{#2}}

\newcommand{\tq}{{\tt q}}
\newcommand\union{\cup}
\newcommand\inter{\cap}
\newcommand\sat{\models}

\newcommand\W{{\cal W}}
\newcommand{\Cert}{{\cal C}}
\newcommand{\Certall}{{\cal C}^+}

\newcommand{\false}{{\it false}}
\newcommand{\true}{{\it true}}
\newcommand{\Aint}{{\tt AIntersect}}

\newcommand{\tr}{{\tt r}}

\newcommand{\nat}{{I\!\!N}}

\newcommand{\validsection}{\mbox{\tt $\la$valid$\ra$}}

\newcommand\tuI{{\tt \tk}} %
\newcommand\tuN{{\tt \tn}} %
\newcommand\tuD{{\tt D}}   %
\newcommand\tuA{{\tt A}}   %
\newcommand\tuS{\tp}       %
\newcommand\tuV{{\tt V}} %
\newcommand\tuR{\mbox{$\tk_r$}}
\newcommand\tuRp{\mbox{$\tk_r'$}}

\newcommand{\Tuples}[1]{{\tt Tuples}(#1)}
\newcommand\As{{\cal S}} %

\newcommand\aisct{{\it Aint}} %

\newcommand\Act{{\it Act}} %
\newcommand\act{{\tt a}} %
\newcommand\Ae{{\cal A}} %

\newcommand\aint{\alpha_{\Ae}} %
\newcommand{\deriv}{\longrightarrow_0^*}
\newcommand{\derivp}{\longrightarrow_1^*}
\newcommand{\derivq}{\longrightarrow_2^*}
\newcommand{\derivi}{\longrightarrow_i^*}
\newcommand{\pdel}{{\it Del}}
\newcommand{\pperm}{{\it Perm}}
\newcommand{\lspki}{{\cal L}_{\it SPKI}}

\newcommand{\Time}{\nat}
\newcommand\run{r}
\def\int{\pi}
\newcommand{\tlna}{L}

\newcommand{\tperm}{{\it P}}

\newcommand{\point}{r,\int,\tk,\con{t}}
\newcommand{\con}[1]{{\tt #1}}
\newcommand{\form}[1]{\phi_{#1}}

\newcommand\now{{\tt now}}

\newcommand\valid{{\it applic}}

\newenvironment{oldthm}[1]{\par\noindent{\bf Theorem #1:} \em
\noindent}{\par}
\newcommand{\othm}[1]{\begin{oldthm}{\ref{#1}}}
\newcommand{\eothm}{\end{oldthm} \medskip}
\newenvironment{oldpro}[1]{\par\noindent{\bf Proposition #1:} \em
\noindent}{\par}
\newcommand{\opro}[1]{\begin{oldpro}{\ref{#1}}}
\newcommand{\eopro}{\end{oldpro} \medskip}

\newcommand{\V}{{\cal V}} 
\newcommand\tuW{{\tt W}} 
\newcommand\tuWt{{\tt W(t)}} 
\newcommand{\Cl}{{\it Cl}}
\newcommand{\Ir}{{\cal I}_r}
\newcommand{\respc}{resp.,\ }

\begin{document}

\title{A logical reconstruction of SPKI%
\thanks{A preliminary version of this appeared in the {\em Proceedings of the
14th IEEE Computer Security Foundations Workshop}, 2001, pp.~59--70.
}}
\author{Joseph~Y. Halpern%
\thanks{Supported in part by NSF under
grants IRI-96-25901 and IIS-0090145, by  ONR under 
grants N00014-00-1-0341, N00014-01-1-0511, and N00014-02-1-0455,
by the DoD Multidisciplinary University Research
Initiative (MURI) program administered by the ONR under
grants N00014-97-0505 and N00014-01-1-0795, 
by AFOSR under grant F49620-02-1-0101,
and by a Guggenheim Fellowship and a
Fulbright Fellowship. Sabbatical support from CWI and the Hebrew
University of Jerusalem is also gratefully acknowledged.
}\\
Cornell University\\
Dept. of Computer Science\\
Ithaca, NY 14853\\
\verb=halpern@cs.cornell.edu=\\
\verb=http://www.cs.cornell.edu/home/halpern=\\
\and
Ron van der Meyden\\
University of New South Wales\\
Australia\\
\verb+meyden@cse.unsw.edu.au+\\
\verb+http://www.cse.unsw.edu.au/+$\sim$\verb+meyden+\\
}
\date{ }
\maketitle

\begin{abstract}
SPKI/SDSI is a proposed public key infrastructure standard that
incorporates the SDSI public key infrastructure.
SDSI's key innovation was the use of {\em local names}.
We previously introduced a Logic of Local Name Containment 
that has a clear semantics and was shown to completely characterize
SDSI name resolution.  
Here
we show how our earlier approach can be extended to deal with a number
of key features of SPKI, including revocation, expiry dates, and tuple
reduction. 
We show that these extensions add 
relatively little complexity to the logic.
In particular, we do not need a nonmonotonic logic to capture
revocation.
We then use our semantics to examine SPKI's tuple reduction rules.
Our analysis highlights places where SPKI's informal description of
tuple reduction is somewhat vague, and shows that extra reduction rules
are necessary in order to capture general information about binding and
authorization.  
\end{abstract}

\section{Introduction}

Rivest and Lampson \citeyear{RL96} introduced SDSI---a Simple
Distributed Security Infrastructure---to facilitate the construction of
secure systems.  In SDSI, principals (agents) are identified with public
keys.  In addition to principals, SDSI allows other names, such as {\tt
poker-buddies}.  Rather than having a global
name space, these names are interpreted {\em locally\/}, by each
principal.  That is, each principal associates with each name a set of
principals.  Of course, the interpretation of a name such as {\tt
poker-buddies} may be different for each agent.  However, a principal
can ``export'' his bindings to other principals using signed
certificates.  Thus, Ron may receive  a signed certificate from the
principal he names {\tt Joe}
describing a set of principals Joe associates with {\tt poker-buddies}.
Ron may then refer to this set of principals by the expression
{\tt Joe's poker-buddies}.  
Rivest and Lampson \citeyear{RL96} give an operational account of local
names; they provide a name-resolution algorithm that, given a
principal $\tk$ and a name $\tn$, computes the set of principals
associated with $\tn$ according to $\tk$.
In \cite{HM01}, building on earlier work of Abadi \citeyear{Abadi98}, we
give a logic LLNC, the Logic of Local Name Containment,
with clean semantics that precisely captures SDSI's operational
name resolution algorithm.  

However, our earlier work made a number of simplifying assumptions to
bring out what we saw as the main issues of name spaces.  In particular,
we (along with Abadi) assumed that certificates never expired and were
not revoked.  SDSI has 
been incorporated into SPKI
\cite{spki1,spki2}, 
which allows expiry dates for certificates
and revocation, and deals with authorization and delegation in addition
to naming.  
In this paper, we show how our earlier approach can be
extended to deal with these features of SPKI.

By not having expiry dates and not allowing revocation, we get a
{\em monotonicity\/} property: having more certificates 
can never mean that fewer keys are bound to a given name.
Heavy use seems to be made of monotonicity in our earlier work.%
\footnote{Ninghui Li \citeyear{Li00} has erroneously
claimed that LLNC is nonmonotonic.
See \cite{HM01} for a rebuttal and discussion of this claim.}
A number of authors have developed 
logical 
accounts of authorization based 
on {\em nonmonotonic\/} logics 
\cite{WL93,JSS97,LGF99,LGF01}.
These are logics where conclusions can be retracted in the presence of
more information (so that $C$ may follow from $A$ but not from $A \land B$).
It has been suggested that revocation
should be modeled using a nonmonotonic logic \cite{LGF99}.%
\footnote{Although it does not go into details about the nonmonotonic
features,  
\cite{LGF99} mentions a logic DL, whose notable features are said to include 
  ``The ability to handle non-monotonic policies. These are policies that deal explicitly 
     with `negative evidence' and specify types of requests that do not comply. 
     Important examples include hot-lists of `revoked' credentials ....
     `Non-monotonic' here means in the sense of logic-based knowledge representation (KR).''
     Some of the authors of this paper have also taken an alternate position: 
     Li and Feigenbaum \citeyear{LF01} recommend that ``a PKI should provide an interface that 
     is monotonic''.
} 
Dealing with nonmonotonicity adds significant complications to a
logic, both conceptually and from a complexity-theoretic point of view
(see, for example, \cite{CaL94}).  Thus, there is a benefit to using a
monotonic logic where possible.  We show that there is no difficulty
capturing expiry dates and revocation (at least as they appear in
SPKI) using a monotonic logic.
Interestingly, 
SPKI's semantics maintains monotonicity even in the presence of
revocation, in the sense that having more certificates (even more revocation
certificates) still allows us to draw more conclusions about both name
bindings and authorizations.  
(Roughly speaking, this is because, in SPKI, a
certificate is ignored unless it is known {\em not\/} to have been
revoked.  Having a revocation certificate issued by $\tk$ covering a
certain time $\ttt$ ensures that there are no other revocation
certificates
issued by $\tk$ covering $\ttt$, and thus allows us to conclude
that certain certificates have not been revoked at time $\ttt$.  Thus,
by having more revocation certificates, we can draw more conclusions.)

We remark that,
although SPKI is monotonic with respect to adding more certificates,
it is not monotonic with respect to time.  
Keys that are bound to a name at time $t$ may no longer be bound to 
that
name at time $t'>t$. 
(Indeed, this does
not require revocation; it suffices that certificates have intervals of
validity.)   
SPKI gives semantics to certificates by first converting them to tuples,
and then providing tuple reduction rules, which are used to reduce the
tuples to a particularly simple form (corresponding to basic 
name binding and authorization decisions). 
We associate with each SPKI certificate a formula in our logic.  Thus,
we have two ways of giving semantics to SPKI certificates: through
tuple reduction and through the logic.  The focus of this paper is on
examining the connection between these two approaches.
Our analysis highlights places where
SPKI's informal description of tuple reduction is somewhat vague, and
shows that extra reduction rules are necessary in order to capture
general information about binding and authorization.  
Besides clarifying ambiguities, the logic allows for reasoning
about the consequences of certain certifications and general reasoning
about naming and authorization.  (See Section~\ref{sec:discussion} for
further discussion of the potential uses of the logic.)

The rest of this paper is organized as follows.  In the next section, we
briefly describe the syntax of SPKI.  In Section~\ref{sec:reduction}, we
describe SPKI's reduction rules.  Section~\ref{sec:logic} describes the
syntax and semantics of our logic for reasoning about SPKI, which
extends LLNC.  In Section~\ref{sec:verification}, we prove our main
results, 
which involve characterizing the power of SPKI reduction rules in terms
of our logic.  
In Section~\ref{related}, we compare our work to 
several 
other
recent approaches to giving semantics to SPKI. 
We conclude in
Section~\ref{sec:discussion} with 
further discussion of the logic. 

\section{SPKI syntax} \label{sec:spki-syntax}

SPKI views authority as being associated with {\em principals}, which it
identifies with public keys. Instead of global names, SPKI has
incorporated SDSI's notion of {\em local name space}.
In SDSI/SPKI, a local name
such as {\tt Joe} is interpreted with respect to a principal, and its
meaning may vary from principal to principal.  There is no requirement
that a local name refer to a unique principal.  For example, a local
name such as {\tt poker-buddies} may refer to a set of principals.
Within its name space, a principal may refer to the interpretation of
names in another principal's name space by means of compound
names. For example, a principal may use the expression {\tt
Joe\s poker-buddies} to refer to the principals that the principal he
refers to as {\tt Joe} refers to as {\tt poker-buddies}.
SPKI calls such expressions {\em compound (SDSI) names}, and uses the
syntax
{\tt (name $\tn_1~ \tn_2 \ldots \tn_k$)}, where the $\tn_i$ are local
names 
(called {\em basic SDSI names} in the SPKI document) 
for $i>1$ and $\tn_1$ is either a local name or a key. 
Such an expression
may also be represented as $\tn_1\s \tn_2\s \ldots \tn_k$.  
A {\em fully-qualified\/} name is one where $\tn_1$ is a key
and $\tn_2, \ldots, \tn_k$ are local names.
While, in general, the interpretation of a compound name depends on the
principal (so that the interpretation of  {\tt Joe\s poker-buddies} by
key $\tk_1$ depends on $\tk_1$'s interpretation of {\tt Joe}, and may be
different from $\tk_2$'s interpretation of {\tt Joe} and {\tt Joe's
poker-buddies}), the interpretation of a fully-qualified name is
independent of the principal.

\commentout{
For
technical reasons and conformity with the notation and terminology in
\cite{HM01}, we refer to SPKI's names as {\em principal
expressions} and define their syntax as follows.
We assume that we are given a set $K$ of {\em keys} and a set
$N$ of {\em local names}.\footnote{
Local names are called {\em byte strings} in SPKI's terminology.
SPKI's keys have internal syntactic structure, but this need not concern
us
here. SDSI also has a notion of
{\em global name}, which has been dropped in SPKI.}
We define a {\em principal expression} to be either a key in $K$, a
local name in $N$, or an expression of the form $\tp\s \tq$
where $\tp$ and $\tq$ are principal expressions.
}

SPKI has other ways of identifying principals.  For example,
SPKI principals may also be the hash of a key, a
threshold subject (an expression representing ``any $m$ out of $N$ of
the following subjects'', used to capture requirements for joint
signatures), or the reserved word
``Self'', representing the entity doing the verification.
For simplicity, in this paper, the only principals we consider are those
defined by compound names.  
\commentout{
A principal expression of the form $\tk\s \tq$, where $\tq$ does not
contain any keys, is called
a {\em fully-qualified principal expression} (SPKI's ``fully
qualified name''). Intuitively, whereas the meaning of
a principal expression such as $\tn\s \tq$ may vary from principal
to principal when $\tn$ is a local name, a fully qualified
principal expression has the same interpretation with respect
to all principals.
SPKI seems to require that all principal expressions be principal
expressions.  While we do not make this requirement here, it
is critical to the proof of our completeness results
(see, for example, the statement and proof of
Theorem~\ref{thm:completeness}).
}%

There are two types of certificates in SPKI, {\em naming certificates}, 
{\em authorization certificates}.  SPKI also has {\em certificate
revocation lists\/} (CRLs); for uniformity, we treat these as
certificates as well.
Again, this seems completely consistent with the SPKI treatment.
A naming certificate has the form of a cryptographically signed message
with contents 
\[{\tt
(cert ~(issuer ~(name~\tuI~\tuN)) ~(subject ~\tuS) ~\validsection)},
\]
where $\tk$ is a key (representing the issuer, whose signature should be
on the certificate), $\tn$ is a local name, $\tp$ is a 
fully-qualified SDSI name,%
\footnote{
SPKI also allows $\tuS$ to be an unqualified name 
(that  is, a string of local names), but notes that in
this case it is to be interpreted as $\tuI\s \tuS$,
which is a fully qualified name.
For simplicity,
we insist upon fully qualified names here.}
and
$\validsection$ is an optional section describing validity constraints
on the
certificate.
The $\validsection$ section may describe an interval during which the
certificate is valid, expressed  by means of 
a ``not-before date'' (expressed in the
syntax as {\tt (not-before $\la$date$\ra$)}) and/or a ``not-after date''
(expressed as
{\tt (not-after $\la$date$\ra$)}).  It may also describe a sequence of
``online
test'' expressions, which   
specify that the certificate should be verified 
either
by
checking a {\em certificate revocation list\/} (CRL) (intuitively, a
list of certificates that have been revoked), by checking a revalidation
list (a list of currently valid certificates), or by performing an
online test.  Each of these components is optional.  

In this paper, we assume that the $\validsection$ field contains only
validity
intervals and a key authorized to sign revocation lists relevant to the
certificate;
the treatment for revalidation lists and other
online tests is similar and does not add new subtleties.%
\footnote{SPKI also
allows a certificate to specify a list of locations where the CRL may be
obtained (rather than requiring that the actual CRL be sent), 
and to provide {\tt http} requests to
these locations with extra parameters, but we ignore these components
since they do not interact with the semantic issues we address.}
We represent dates as natural numbers.  From the {\tt not-before}
and {\tt not-after} sections of a certificate we may obtain a validity
interval $\tuV =[\ttt_1,\ttt_2]$ where 
$\ttt_1,\ttt_2 \in \nat\cup \{ \infty\}$ are
respectively the not-before time and the not-after times indicated.
If no {\tt not-before} time is given, we take $\ttt_1 = 0$ and,
similarly, if no {\tt not-after} time is given, we take $\ttt_2 =
\infty$.  We assume that $\ttt_1\leq \ttt_2$, so that the validity
interval
is nonempty. 
We also allow the empty interval, which we denote $\emptyint$.

For simplicity, we abbreviate naming certificates as  
$({\tt cert}~\tuI~\tuN~\tuS~\tuV~\tuR)$,
where $\tuV$ has the form $[t_1,t_2]$ and $\tuR$ is the key 
which is authorized to sign CRLs relevant to the certificate.
The $\tuR$ component may not be present: if it is, then we say that 
the certificate is {\em revocable by $\tuR$}. %
We occasionally write
$({\tt cert}~\tuI~\tuN~\tuS~\tuV~\la\tuR \ra)$ to denote a generic
naming certificate where $\tuR$ may or may not be present.
A naming certificate binds the 
fully-qualified name
$\tuS$ to the local name $\tuN$ in $\tuI$'s local name space during
the period $\tuV$, provided that certificate does not appear in any
CRLs
signed by $\tuR$.  Binding $\tuS$ to $\tuN$
means that the 
interpretation of $\tuN$ with respect to $\tuI$ 
includes the
meaning of $\tuS$.
For example, if Ron and Joe are principals, then the certificate
$({\tt cert}~{\tt Ron}~{\tt doctor}~({\tt Joe's~doctor}) ~[1,3])$ binds
{\tt Joe's~doctor} to the local name {\tt doctor} in {\tt Ron}'s  local name space 
from time 1 to time 3; 
moreover, this
certificate is irrevocable.

Authorization certificates have the form
\[{\tt 
\small 
(cert ~(issuer ~\tuI) ~(subject ~\tuS) 
~(propagate) ~\tuA 
~\validsection)
},
\]
where $\tuI$ is a key, $\tuS$ is a
fully-qualified name,%
\footnote{SPKI also allows the subject to be an 
unqualified name. We make the simplifying assumption of 
qualified names just as we did above.}
$\tuA$ is 
what the SPKI document calls an {\em authorization\/} and we call an
{\em action expression}, since it
 represents a set of actions,
and $\validsection$ is a validity section, as
described above. 
The ``(propagate)'' section is optional.
Intuitively, the issuer uses such a certificate to grant the
subject the authority to perform the actions in $\tuA$.
Moreover, if
``(propagate)'' is present, 
then the subject is further authorized
to propagate this authority to others.%
\footnote{
Note that
if
``(propagate)'' is not present, 
then we treat this as there
being no indication of whether the subject is authorized to propagate
the authority to others, rather than it being the case that the subject
is {\em not\/} permitted to propagate the authority.  This allows it to
be consistent for $\tk$ to issue two certificates that are identical
except that one contains ``(propagate)'' while the other does not.
Under our interpretation, the former supersedes the latter. 
(This seems particularly reasonable if we assume that someone who is
seeking permission to perform an action will present those certificates
that maximize  his/her rights.)
The SPKI document is silent on this issue.}
We abbreviate authorization certificates as
${\tt (cert~\tuI~\tuS~
\tuD~\tuA~
\tuV~\tuR)}$,
where $\tuD$ is 
a
a Boolean (which stands for {\em delegate}) indicating whether or not
propagation is permitted.
Again, the $\tuR$ component is optional, and we use
${\tt (cert~\tuI~\tuS~
\tuD~\tuA~
\tuV~\la \tuR \ra)}$ to denote a generic
authorization certificate where the $\tuR$ component may or may not be
present. 
SPKI takes an action expression $\tuA$ to be an {\em
S-expression\/}---a list of strings or sublists.  It uses $\Aint$ to
denote the intersection of 
action expressions.%
\footnote{Howell and Kotz \cite{HK00a}
have noted some problems with intersection for SPKI's action expressions (or {\em tags}), 
namely that not all intersections of tags can be represented as a tag. The problem 
can be eliminated by extending the set of tags. 
We simply avoid the issue here by treating action expressions very abstractly, 
and assuming that they can always be intersected.} 
As we suggested above, 
action expressions
are best thought of as
sets of actions. 
We abstract this by assuming that there is some set $\Act$ of actions
and a set $\Ae$ of 
action
expressions that, intuitively, represent sets of actions in $\Act$.  We
assume that $\Ae$ includes 
all
finite subsets $\{\act_1, \ldots, \act_n\}$
of $\Act$.  
Moreover, we assume that, given two action
expressions $\tuA_1$ and $\tuA_2$ in $\Ae$, we can
easily compute a third action expression in $\Ae$,
denoted $\tuA_1 \inter \tuA_2$, which intuitively represents the
intersection of the sets represented by $\tuA_1$ and $\tuA_2$.  
(We capture this intuition by a semantic constraint below.) 
The reason that we
allow
expressions in $\Ae$, rather than just finite subsets of $\Act$, is that
SPKI allows (some) 
expressions that represent
infinite sets of actions.  For example, SPKI 
allows an action expression 
of the form ({\tt ftp ftp.clark.net /pub/cme/*}),
which allows access to directories ftp.clark.net that start with
/pub/cme/ (see \cite[Section 6.3.1]{spki1}).

We assume that there is a fixed {\em action interpretation\/} $\aint$ that maps
action expressions in $\Ae$ to subsets of $\Act$. 
We require that 
if $\tuA$ is a finite
subset of $\Act$, then 
$\aint(\tuA) = \tuA$, and that
$\aint(\tuA_1 \inter \tuA_2) = \aint(\tuA_1) \inter
\aint(\tuA_2)$.   
We also assume that we can decide (given $\tuA_1$ and $\tuA_2$) 
whether $\aint(\tuA_1 \inter \tuA_2) = \aint(\tuA_2)$
(intuitively, whether $\tuA_1$ denotes a subset of $\tuA_2$).

\commentout{
The authorization $\tuA$ in a SPKI certificate is
required to be an ``s-expression'', a type of syntactic structure used
throughout the SPKI specification. SPKI views the semantics of
authorizations to be a matter to be determined by each application
built using the SPKI infrastructure. However, SPKI does specify an
operation $\Aint$ used to {\em intersect} two authorizations, and any
application-defined semantics needs to be consistent with this
operation.  The details of the s-expression syntax and the definition
of $\Aint$ need not concern us here, as our treatment of
authorizations will be more abstract than SPKI's. It suffices to note
that, intuitively, authorizations represent sets of actions, and that
$\Aint(\ae_1,\ae_2)$ is intended to be an s-expression representing
the intersection of the sets represented by $\ae_1$ and $\ae_2$.

We abstract SPKI's authorizations as follows.
An {\em authorization space} is a tuple $\As = \la \Ae, \aisct, \Act,
\aint\ra$
where
\be
\item $\Ae$ is a set of {\em authorization expressions}, containing
$\top$ and $\bot$,
\item $\aisct: \Ae \times \Ae \rightarrow \Ae$ is an operation
on authorization expressions,
\item $\Act$ is a set of {\em actions},
\item $\aint: \Ae \rightarrow \pow{\Act}$ is a
function interpreting authorization expressions as sets of actions
\ee
Intuitively, the set $\Ae$ of authorization expressions is intended to
be a set of syntactic expressions corresponding to SPKI's authorization
s-expressions. We require that $\aint(\aisct(\ae,\ae')) = \aint(\ae)
\cap \aint(\ae')$ for all $\ae, \ae' \in \Ae$. That is, the syntactic
operation on action expressions corresponds at the semantic level to
intersection of the sets of actions represented. Additionally,
we require that $\aint(\top) = \Act$ and $\aint(\bot)= \emptyset$.
We assume some fixed authorization space for the remainder of the
paper.
}%

A CRL has the form 
\[{\tt 
(crl ~(canceled~c_1, \ldots,c_n)~V)}\]
where the $\tc_i$ are hashes of certificates.%
\footnote{SPKI also allows delta-CRLs. We omit these since they do 
not introduce essentially new semantic issues.}
It is left implicit that the CRL needs to be signed by some key $\tk$.
For simplicity, we will assume that CRLs contain certificates 
themselves rather than hashes. 
We abbreviate a CRL as 
$({\tt crl~ \tuI~ (canceled~c_1, \ldots,c_n)~V}).$
Intuitively, this says that, according to the issuer $\tuI$, the
certificates ${\tt c_1, \ldots, c_n}$ are revoked during the interval
$\tuV$.
We require that each of the certificates $\tc_i$ be revocable by $\tuI$
(otherwise  $\tuI$ is attempting to revoke a certificate 
that it is not entitled to revoke). 

Let $\Certall(K,N,\Ae)$ consist of all certificates 
over $(K,N,\Ae)$ (i.e., where all the keys are in $K$, all the names
used are in $N$, and all the 
action
expressions are in $\Ae$);
let 
$\Cert(K,N,\Ae)$ be the subset of $\Certall(K,N,\Ae)$ consisting of
all naming and authorization certificates; 
and let $\Cert_R(K,N,\Ae)$ be the subset of $\Certall(K,N,\Ae)$
consisting of all CRLs.

\section{SPKI's tuple reduction rules}\label{sec:reduction}

The semantics of SPKI certificates is characterized by a description
of the algorithm invoked to verify that a sequence of credentials
supports an authorization decision \cite[Section 6]{spki1}. 
It is left to the prover (the
agent presenting a set of credentials) to construct an appropriate
sequence before submitting a request to the verifier.

Given a set $C$ of naming and authorization certificates and a
set $C_R$
of 
CRLs,
the algorithm first converts these
certificates to a set of {\em tuples}, and reduces these tuples
according to certain rules.  

There are two types of tuples:
4-tuples, related to name binding certificates, and
5-tuples, related to  authorization certificates.
A 4-tuple has the form $\la \tuI,\tuN,\tuS,\tuV \ra $ where the
components are exactly as in the first four components of a naming
certificate.  Similarly, a 5-tuple has the form $\la
\tuI,\tuS,
\tuD,\tuA,
\tuV \ra $, where the 
components are exactly as in 
an authorization certificate.
Note that neither the 4-tuples nor the
5-tuples mention the $\tuR$ component of 
authorization certificates.
We use $\tau_{\tc}$ to denote the 4- or 5-tuple corresponding to
certificate $\tc$.

The first step in the conversion is to check each certificate in $C$ 
to see if it has been revoked.  
Given a naming certificate $\tc =
({\tt cert}~\tuI~\tuN~\tuS~[\ttt_0,\ttt_1]~\tuR)$ and a 
CRL
$\tc_R={\tt (crl~\tuRp~(
canceled
~c_1, \ldots,c_n)~[\ttt_0',\ttt_1'])}$,
say that $\tc$ is {\em live\/} with respect to $\tc_R$ if 
\begin{enumerate}
\item $\tc$ is signed by $\tuI$, and
\item 
the following four conditions all hold:
\begin{enumerate}
\item $\tuR = \tuRp$, 
\item $\tc_R$ is signed by $\tuRp$, 
\item $[\ttt_0,\ttt_1] \inter [\ttt_0', \ttt_1'] \ne \emptyint$,
\item $\tc \notin \{\tc_1, \ldots, \tc_n\}$.
\end{enumerate}
\end{enumerate}
Intuitively, $\tc$ is live with respect to $\tc_R$ if $\tc$ is properly
signed,
the validity component in $\tc$ requires checking a CRL, $\tc_R$ is
the certificate appropriate for the CRL and, according to the CRL, $\tc$
has not been revoked.  
If $\tc$ is live with respect to $\tc_R$, define
$\tau(\tc,\tc_R)$ to be the 4-tuple 
$(\tuI~\tuN~\tuS~[\ttt_0,\ttt_1] \inter [\ttt_0',\ttt_1'])$. 
If $\tc$ is an authorization certificate, there is an
essentially identical notion of liveness with respect to $\tc_R$ and
corresponding 5-tuples $\tau(\tc,\tc_R)$.
We leave details to the
reader.
Define $\Tuples{C,C_R}$ to be the set of tuples
$\tau(\tc,\tc_R)$ where $\tc\in C$, $\tc_R\in C_R$, and $\tc$ is live
with respect
to $\tc_R$, 
together with the set of tuples 
$\tau_{\tc}$
where $\tc\in C$ is irrevocable.

The mapping $\tau(\tc,\tc_R)$ is our attempt to capture the mapping
described in \cite[Section 6]{spki1}, which says:
\begin{quote}
Individual certificates are verified by checking their
signatures and possibly performing other work.  They are then
 mapped to intermediate forms, called ``tuples'' here.
The other work for SPKI or SDSI certificates might include
processing of on-line test results (CRL, re-validation or one-time
validation). \ldots \ If on-line tests are involved in the certificate
processing, then the validity dates of those on-line test results are
intersected \ldots with the validity dates of the certificate to yield
the dates in the certificate's tuple(s).
\end{quote}

Note that the mapping $\Tuples{C,C_R}$ is
monotonic: if $C' \supseteq C$, $C_R' \supseteq C_R$, 
then $\Tuples{C',C_R'}\supseteq
\Tuples{C,C_R}$. Intuitively, a certificate $\tc \in C$ that is
revocable by $\tuR$ is 
considered to be valid at time $\ttt$ if there is clear evidence that $\tc$ has
not been revoked by $\tuR$ at time $\ttt$, where the ``evidence'' is
that there is a CRL issued by $\tuR$ that covers time $\ttt$ that does
not mention $\tc$.  The absence of a certificate $\tc$ from
{\em any} relevant CRL $\tc'$ ensures that the statement being made by
that certificate applies during the intersection of the intervals of
$\tc$ and $\tc'$. 

The intuition that 
$\Tuples{C,C_R}$ 
consists of the still valid certificates
does not hold up so well when there can be more than one CRL
relevant to the validity of a certificate $\tc$ at a time $\ttt$.
For suppose that $\tc$ is revoked according to one certificate and not
revoked according to another.
One could reasonably argue
that, in this situation, the two CRLs are in conflict about whether
the certificate has been revoked, and either could apply.  In
particular, the outcome of an authorization decision would depend on
which CRL is presented in support of a request.  To avoid such
nondeterminism, SPKI \cite[Section 5.2]{spki1} assumes that at any
time, at most one CRL applies.  We say a set $C_R$ of CRLs is {\em
consistent\/} if 
it is not the case that there exist CRLs $\tc, \tc'
\in C_R$, both issued by $\tk$, with validity periods $\tuV,\, \tuV'$,
respectively, such that $\tuV\cap \tuV' \ne \emptyset$.
Restricting to
consistent sets of CRLs ensures that it is safe to take any relevant
CRL not containing a certificate as evidence for the validity of that
certificate, since there cannot exist a CRL contradicting this
conclusion.  
This assumption supports the monotonicity of
$\Tuples{C,C_R}$, and is essentially what allows us to use a monotonic
logic, even in the presence of revocation.%
\footnote{Of course, in practice, it may well be that a set of CRLs is
inconsistent.  Both the SPKI document and our paper are silent on what
to do in this case.  
}

The semantics of SPKI given in \cite{spki1} is in terms of tuple
reduction.
However, the presentation of how the tuples are intended to be
used to make an authorization decision is not completely formal.
\cite[Section 6]{spki} states that ``Uses of names are replaced with
simple
definitions (keys ...), based on the name definitions available from
reducing name 
4-tuples''
and that ``Authorization 5-tuples are then
reduced to a final authorization decision''.
The rule for 5-tuple reduction required for the latter 
step is explicitly described (in \cite[Section 6.3]{spki1}); 
it combines two 5-tuples to produce another 5-tuple: 
\be
\item[R1.] $\la \tk_1,
\tk_2, 
\true, \tuA_1, \tuV_1\ra +
         \la 
\tk_2,
\tuS, \tuD_2, \tuA_2, \tuV_2\ra
       \longrightarrow
       \la \tk_1,\tuS, \tuD_2, \tuA_1 \inter \tuA_2, \tuV_1\cap
\tuV_2\ra$.%
\footnote{SPKI uses ${\tt Vintersect}$ to
denote the intersection of timing expressions; we use the simple
$\inter$ symbol here.  The intersection of timing intervals is defined
in the obvious way.
If $\tuV_1 \inter \tuV_2$ is empty, then $\tuV_1 \inter \tuV_2 =
\emptyset$,
since we are using $\emptyint$ to denote the empty interval.
}  
\ee
Intuitively, if $\tk_1$ permits $\tk_2$ to delegate authority to the
actions in $\tuA_1$ during $\tuV_1$ and $\tk_2$ gives $\tp$ authority
over
the actions in $\tuA_2$ (and to further delegate authority, if $\tuD_2$
is $\true$) for the interval $\tuV_2$, then this is tantamount to
$\tk_1$ giving authority to $\tp$ over the actions in $\tuA_1 \inter
\tuA_2$ (and to further delegate authority, if $\tuD_2$ is $\true$) for
the interval $\tuV_1 \inter \tuV_2$.

The way that 4-tuples are to be reduced is slightly less transparent.
The discussion of 4-tuple 
reduction in \cite[Section 6.4]{spki1}
does not describe rules by which 4-tuples may be reduced, 
but rather shows how 
fully qualified names 
may be rewritten 
using 4-tuples. However, the discussion suggests
the following rule for 4-tuple reduction: 
\be
 \item[R2.] $\la \tk_1,\tn, \tk_2\s \tm\s \tp, \tuV_1\ra +
        \la \tk_2,\tm, \tk_3, \tuV_2\ra         \longrightarrow
        \la \tk_1,\tn, \tk_3\s \tp, \tuV_1 \cap \tuV_2\ra$.
\ee
We allow $\tp$ to be the empty string in this rule, treating 
an expression of the form 
$\tr\s \tp$
as equal to 
$\tr$ 
in this case, to avoid the need for stating the rule that
results from replacing $\tk_2\s \tm\s \tp$ by $\tk_2\s \tm$ and
replacing $\tk_3\s \tp$ by $\tk_3$.  We use this convention in stating
other 
rules too.  
Intuitively, this rule says that if $\tk_1\s \tn$ is bound
to 
$\tk_2\s \tm\s \tp$ for the interval $\tuV_1$, and
$\tk_2\s \tm$ is bound to $\tk_3$ for the interval $\tuV_2$,
then $\tk_1\s \tn$ will be 
bound to $\tk_3\s \tp$ for the interval $\tuV_1 \inter \tuV_2$.  
The SPKI document also considers a generalization of this rule:
\be
 \item[R2$'$.] $\la \tk_1,\tn, \tk_2\s \tm\s \tp, \tuV_1\ra +
        \la \tk_2,\tm, \tq, \tuV_2\ra         \longrightarrow
        \la \tk_1,\tn, \tq\s \tp, \tuV_1 \cap \tuV_2\ra$. 
\ee
We consider in Section~\ref{sec:verification} the role of R2 vs.~R2$'$.

The step of the authorization decision process described as ``Uses of
names are replaced with simple definitions (keys ...), based on the
name definitions available from reducing name 4-tuples'' is not
further formalized.  
\commentout{
We propose here that it may be captured by
3-tuples of the form $\la \tp,\tq,\tuV\ra$, where $\tp$ and $\tq$ are
fully
qualified principal expressions and $\tuV$ is a validity interval.
Intuitively, a 3-tuple of the form $\la \tp, \tq, \tuV\ra$ says that the
name $\tp$ is bound to the name $\tq$ during the interval $\tuV$.
We have 
the following rule for 3-tuple reduction: 
\be 
\item[] $\la \tp,\tk\s \tn\s \tq, \tuV_1\ra + \la \tk, \tn,
\tr,\tuV_2\ra
\longrightarrow \la \tp, \tr\s \tq, \tuV_1\cap \tuV_2\ra $.
\ee
These rules capture 
the intent of the rules for name reduction 
described in \cite[Section 6.4]{spki1}.  
The introduction 
of 3-tuples enables us to model the 
replacement of names by keys by the rule 
\be 
\item[] 
$\la \tk_1,\tp, \tuD_1, \tuA_1, \tuV_1\ra +
         \la \tp, \tk_2, \tuV_2\ra
       \longrightarrow
       \la \tk_1,\tk_2, \tuD_1, \tuA_1, \tuV_1\cap \tuV_2\ra$,
\ee
provided that $\tuA_1, \tuV_1\cap \tuV_2\ra$ is nonempty.

[[ As I suggested in my email, I think we can avoid the use of
3-tuples and use:

$\la \tk_1,\tp, \tuD_1, \tuA_1, \tuV_1\ra +
         \la \tk_1, \tp, \tq, \tuV_2\ra
       \longrightarrow
       \la \tk_1,\tq, \tuD_1, \tuA_1, \tuV_1\cap \tuV_2\ra$.
}%
However, the following rule seems to capture this intuition:

\be \item[R3.] $\la \tk_1,\tk_2\s \tn\s \tp, \tuD, \tuA, \tuV_1\ra +
         \la 
\tk_2, 
\tn, \tk_3, \tuV_2\ra
       \longrightarrow
       \la \tk_1,\tk_3\s \tp, \tuD, \tuA, \tuV_1\cap \tuV_2\ra$.
\ee

Again, it is possible to generalize R3 much the same way as R2$'$
generalizes R2.
\be \item[R3$'$.] $\la \tk_1,\tk_2\s \tn\s \tp, \tuD, \tuA, \tuV_1\ra +
         \la \tk_2, \tn, \tq, \tuV_2\ra
       \longrightarrow
       \la \tk_1,\tq\s \tp, \tuD, \tuA, \tuV_1\cap \tuV_2\ra$.
\ee

As we shall see, the question of whether we use R2/R3 or R2$'$/R3$'$ has a
nontrivial impact on the type of conclusions we can draw using tuple
reduction; see, for example, Theorems~\ref{thm:concretecompleteness}
and~\ref{thm:complete}.

\commentout{
I believe that use of this rule allows the
elimination of the 3-tuples, but it is arguable how well it captures
the sense of ``replace names by keys''.  While this is OK for
deciding whether a key is authorized to do something, it leads to
incompleteness for more general questions, such as ``are all keys in
the meaning of a given expression authorized to do A?''. 
Suppose that it can be proved that $\now \in
V_2 \rimp \tp \bdto \tq$, where $\tq$ is longer than $\tp$.  Then we
want to be able to reduce $\la \tk_1,\tp, \tuD_1, \tuA_1, \tuV_1\ra$
to $\la \tk_1,\tq, \tuD_1, \tuA_1, \tuV_1\cap \tuV_2\ra$. The rule above

does not allows this, since it always decreases the size of the
principal 
expression. 
It seems that a completeness result is possible with a 
slight generalization: either 
\be
\item[] $\la \tk_1,\tk_2\s \tn\s \tp, \tuD_1, \tuA_1, \tuV_1\ra +
         \la \tk_2, \tn, \tq, \tuV_2\ra
       \longrightarrow
       \la \tk_1,\tq\s \tp, \tuD_1, \tuA_1, \tuV_1\cap \tuV_2\ra$
\ee
or, using 3-tuples, 
\be
\item[] $\la \tk_1,\tp, \tuD_1, \tuA_1, \tuV_1\ra +
         \la \tp, \tq, \tuV_2\ra
       \longrightarrow
       \la \tk_1,\tq, \tuD_1, \tuA_1, \tuV_1\cap \tuV_2\ra$
\ee
The three tuples are probably 
useful as part of the completeness result in any case. 
]]

}%

\commentout{
If $T$ is a set of tuples and $\tau$ is a tuple, we write
$T \deriv \tau$ if there exists a sequence of tuples
$\tau_1, \ldots , \tau_k$ such that $\tau_k = \tau$ and for each $i\leq
k$
either $\tau_i \in T$ or there exist $j,j'<i$ such that
$\tau_j + \tau_{j'} \longrightarrow \tau_i$ is an instance of
R1--R3.}

SPKI intends that the tuple reduction rules play several different
roles. Besides being used for specific authorization decisions, the
tuple reduction process is intended to serve as a means of
derivation of consequences of a set of certificates that can form the
basis of a ``certificate result certificate'' that captures a set of
authorizations that may be derived from a collection of certificates
\cite[Section 6.6]{spki1}.  
As we shall see, in a precise sense, R1--3 suffice for making specific
authorization decisions
at a given time.
However, to derive 
more general
consequences of a
set of certificates, we need to consider R2$'$ and R3$'$, as well as
other rules discussed in Section~\ref{sec:verification}.
In the next section, we provide a semantics for SPKI that lets us
provide semantics for the reduction rules; we then
use that semantics in Section~\ref{sec:verification} to carefully
examine the rules.

\section{A logic for reasoning about SPKI}\label{sec:logic}

We now define a formal language $\lspki(K,N,\Ae)$ for reasoning about 
SPKI. $\lspki(K,N,\Ae)$ is an extension of the language LLNC defined in
\cite{HM01}. The parameters $N$ and $\Ae$ (the set of names and the set
of
action expressions) do not play a significant role.  However, for
some of our results, the cardinality of the set $K$ does play a role.
To simplify the notation, we often omit the parameters 
that play no significant role,
and write, for example,  $\lspki$ or $\lspki(K)$.
We do the same in all other contexts where these parameters are used.

\subsection{Syntax}
Following \cite{HM01}, given a set $K$ of keys and a set $N$ of local
names, we define a {\em principal expression} (over $K$ and $N$) to be
either a key in $K$, a local name in $N$, or an expression of the form
$\tp\s \tq$ where $\tp$ and $\tq$ are principal expressions.  The
compound names of SPKI/SDSI can be viewed as principal expressions.
Note that parenthesization matters for principal expressions; for
example, $(\tn_1\s \tn_2)\s \tn_3$ is different from $\tn_1\s (\tn_2\s
\tn_3)$.  However, our semantics guarantees that the combination of
names is associative, so that, in fact, the two principal expressions
are equivalent (see Lemma~\ref{lem:associative}.  For definiteness,
we assume that all principal expressions that
arise in naming and authorization 
certificates
are parenthesized to 
the right.%
\footnote{Recall that, in SPKI, names are just written as $({\tt
name~n_1~
\ldots~n_k})$, so there is no parenthesization involved at all.}

The primitives of $\lspki(K,N,\Ae)$ consist of 
\begin{itemize}
\item principal expressions 
over $K$ and $N$;
\item the set $\Certall(K,N,\Ae)$ of naming, authorization
and revocation certificates
that can be formed from $K$, $N$, and $\Ae$;

\item a special constant $\now$; 
\item {\em validity intervals\/} $\tuV$ consisting of pairs $[\ttt_1, \ttt_2]$
of 
times in $\Time \union \{\infty\}$  with $\ttt_1 \le \ttt_2$,
together with the empty interval $\emptyint$.
\end{itemize}

The set of formulas of $\lspki(K,N,\Ae)$ is the smallest set such that
\be
\item if $\tp$ and $\tq$ are principal expressions then $\tp \bdto \tq$
is a formula; 

\item if $\tc \in \Cert(K,N,\Ae)$, then $\tc$ and 
$\valid(\tc)$ are formulas;

\item if $\tk$ is a  key, $\tp$ is a principal expression, 
and  $A \in \Ae$, then $\pperm(\tk,\tp,A)$ and
$\pdel(\tk,\tp,A)$ are formulas;

\item $\now \in V$ is a formula;

\item if $\phi,\psi$ are formulas then $\neg \phi$ and $\phi \land \psi$
are formulas.
\ee

Intuitively, $\tp \bdto \tq$ says that all the keys in $\tq$ are bound
to $\tp$.  Since principal expressions are associated with sets of keys,
this just says that the keys associated with $\tq$ are a subset of those
associated with $\tp$.  The formula $\tc$ is true at a time $\ttt$ if
the certificate $\tc$ was issued before $\ttt$.  (To make sense of this,
the semantic object which determines whether formulas are true must
include a list of the certificates that have been issued and the current
time.)  The formula $\valid(\tc)$ is true 
at a time $\ttt$ if $\tc$ is either irrevocable or if it is revocable,
but is known not to have been revoked at time $\ttt$.  Although we read
$\valid(\tc)$ as ``$\tc$ is applicable'', it is worth noting that
$\valid(\tc)$ could be true at time $\ttt$ even if $\tc$ was not issued
before time $\ttt$ or its validity interval does not include $\ttt$.
There are other formulas in the logic that enable us to say that $\tc$
has been issued (namely, the formula $\tc$) and that the current time is
in a given validity interval (namely, $\now \in V$).  Finally, as the
notation suggests, the formula $\pperm(\tk,\tp,A)$ says that $\tk$
permits $\tp$ to perform the actions in $A$ and $\pdel(\tk,\tp,A)$ says
that $\tk$ permits $\tp$ to delegate authority over the actions in $A$.

LLNC can be viewed as the fragment of $\lspki$ where the only
certificates allowed are those of the form $({\tt cert}~ \tk~ \tn~
\tp)$,
which corresponds to the LLNC formula $\tk \cert \tn \bdto \tp$.  
There are no 
formulas in LLNC of the form $\now \in \tuV$, 
$\pperm(\tk,\tp,A)$, $\pdel(\tk,\tp,A)$, 
or $\valid(\tc)$
(since there is no notion of
time in LLNC, and permission, delegation, and revocation are not
treated).%
\footnote{LLNC does allow formulas of the form $\tk \cert
\phi$ for arbitrary formulas $\phi$.  However, if $\phi$ is not of the
form $\tn \bdto \tp$, then such formulas 
do not interact with the other constructs
under the semantics of \cite{HM01}.  
Thus, LLNC does not gain additional expressive power from such formulas.

Following SDSI, LLNC also  has a notion of a {\em
global name}.  Since global names have been omitted in SPKI, we omit
them in $\lspki$ as well.}

\subsection{Semantics}
\label{sec:minmod}

The semantics for $\lspki$ extends that of LLNC.  
We begin by outlining the main components of the semantic model.

In LLNC, there is a notion of a {\em world}.
A world
essentially describes which certificates have been issued.  Since now we
have time in the picture, we need a temporal analogue of a world.  This
is a run.  Formally, 
a $(K,N,\Ae)$-{\em run} is a function $\run:\Time \rightarrow
\pow{\Certall(K,N,\Ae)}$. 
(We use $\pow{X}$ to denote the set of subsets of $X$ here and
elsewhere). 
We are implicitly assuming a 
global clock and are taking time with respect to that global clock.
Intuitively, $r(\tts)$ is the set of appropriately signed certificates
issued 
at time $\tts$. That is, if $\tc$ is a certificate, 
then $\tc \in r(\tts)$ 
if a 
certificate with contents $\tc$ 
is issued at time 
$\tts$
in run $r$.
For compatibility with the SPKI document, we assume that the set of
revocation certificates issued in $r$ (that is, the set of revocation
certificates in 
$\union_{\tts\in \Time} r(\tts)$) 
is consistent: there cannot be
two CRLs with the same issuer whose validity intervals overlap.
As we said before, we assume that there is a set $\Ae$ of 
action
expressions, which represent sets of actions in a set $\Act$,  
and a fixed action interpretation $\aint$ that maps expressions in $\Ae$
to subsets of $\Act$.

To interpret local names, LLNC has a construct called a 
{\em local name assignment\/} that associates with each key $\tk$ and
local
name $\tn$ the set of keys bound to $\tn$ by $\tk$.  There is an
analogous function
here, but it now takes a time as an argument, since
the association may vary over time.  In addition, to take into account
the new constructs in $\lspki$, there is 
a function that
associates with each key $\tk$, local name $\tn$, and time $\tts$ the
set
of actions that $\tk$ has granted each other principal permission to
perform, and describes whether or not that permission can be delegated.
These functions can be extended from local names to all principal
expressions; see below.

Formally,
\bi
\item
a {\em (temporal) local name assignment} 
(for $K$ and $N$)
is a function $\tlna:
K\times N\times \nat \rightarrow \pow{K}$.  Intuitively, for $\tk \in
K$, $\tn \in N$ and $\con{t} \in \nat$, the set $\tlna(\tk,\tn,\con{t})$
contains the
keys associated at time $\con{t}$ with the name $\tn$ in $\tk$'s 
name space.

\item
a {\em (temporal) permission/delegation assignment} 
(for $K$ and $\Act$)
is a function
$\tperm: K\times \nat \times  K\times \Act \rightarrow \{0,1,2\}$
such that 
if $\tperm(\tk_1,\con{t},\tk_2,\act) = 2$ and 
$\tperm(\tk_2,\con{t},\tk_3,\act) = i$, then 
$\tperm(\tk_1,\con{t},\tk_3,\act) \ge i$.
Intuitively, $\tperm(\tk,\con{t},\tk',\act)$ is 
0 if at time $\con{t}$, $\tk$ has not granted $\tk'$ the right to
perform or
delegate $\act$; it is 1 if  principal $\tk$ 
has granted permission to principal $\tk'$ to perform action $\act$;
it is 2 if, in addition, principal $\tk$ has delegated 
authority to principal $\tk'$ to propagate the right to perform action
$\act$.%
\footnote{As noted in \cite{spki}, there is not much point to having a
principal able to propagate the right to perform an action without
having
the right to perform it, since the principal may always grant
itself that right.}
The meaning of the right to propagate is captured by the 
condition above: 
if, at time $\con{t}$, $\tk_1$ has granted $\tk_2$ the right to
propagate  permission to perform $\act$, and $\tk_2$ has granted
permission to perform (propagate) $\act$, then, according to $\tk_1$, 
principal $\tk_3$ has the right 
to
perform (propagate) $\act$.
The reason that the condition says $P(\tk_1,\ttt,\tk_3,
\act
) \ge i$
rather than $P(\tk_1,\ttt,\tk_3,\act) = i$ is that if $i = 1$, for
example, it is possible that $\tk_1$ independently granted $\tk_3$
the right to delegate $\act$, so that $P(\tk_1,\ttt,\tk_3,\act) = 2$.
\ei
A 
$(K,N,\Act)$ 
{\em interpretation\/} $\pi$ 
is a tuple $\la L, P \ra$
consisting 
of a local name assignment $\tlna$ 
for $K$ and $N$ and
a permission/delegation assignment $\tperm$
for $K$ and $\Act$.
We omit the modifier $(K,N,\Act)$ 
when it is not relevant to the
discussion.  However, it is important to note that the parameters that
characterize the language also characterize the interpretations.

Given 
a local name assignment $L$,
a key
$\tk$, and a time $\con{t}\in \Time$,
we can assign to each principal expression $\tp$ 
a set of keys
$\intension{\tp}{L,\tk,\con{t}}$.
This set is defined by the following recursion:
\bi
\item $\intension{\tk'}{L,\tk,
\ttt} = \{\tk'\}$, if $\tk'\in K$ is a key,

\item $\intension{\tn}{L,\tk,\con{t}} = L(\tk,\tn,\con{t})$, if $\tn\in
N$ is a local name,

\item $\intension{\tp\s \tq}{L,\tk,\con{t}} = \bigcup \{
\intension{\tq}{L,\tk',\con{t}} ~|~ \tk'\in
\intension{\tp}{L,\tk,\con{t}} \}$.
\ei

This definition is essentially identical to that in \cite{Abadi98,HM01},
except that we have made the interpretation of local names depend on
the time of evaluation.

It is now easy to prove some basic facts about principal expressions.
First, we can show that a fully-qualified name is independent of the
key.
\lem If $\tp$ is a fully qualified principal expression, then 
$\intension{\tp}{L,\tk,\tts} = \intension{\tp}{L,\tk',\tts}$ for all
keys $\tk$ and $\tk'$. \elem

\prf By an easy induction on the structure of $\tp$. \eprf

We also make precise our claim that the combination of names is
associative.
\lem\label{lem:associative} For all principal expressions $\tp_1$,
$\tp_2$, and $\tp_3$, keys $\tk$, and local name assignment $L$, 
$$\intension{\tp_1\s(\tp_2\s \tp_3)}{L,\tk,\tts} = \intension{(\tp_1\s
\tp_2)\s \tp_3}{L,\tk,\tts}.$$
\elem
\prf By unwinding the definitions, it immediately follows that both
$\intension{\tp_1\s(\tp_2\s \tp_3)}{L,\tk,\tts}$ and $\intension{(\tp_1\s
\tp_2)\s \tp_3}{L,\tk,\tts}$ are equal to
$$\union \{ \intension{\tp_3}{L,\tk_2,\tts}: \tk_2 \in
\intension{\tp_2}{L,\tk_1,\tts}, \tk_1 \in \intension{\tp_1}{L,\tk,\tts}
\}.$$ 
\eprf

In order to capture the impact that CRLs have on the
interpretation of naming and authorization certificates, we say that a
certificate $\tc$ is {\em applicable} at time $\tts$ 
in a run $r$ 
if either $\tc$ is not
revocable, or $\tc$ is revocable by a key $\tuR$ and for some $\tts'
\leq \tts$ we have 
$({\tt crl~ \tuR~ (canceled~c_1, \ldots,c_n)~V}) \in
r(\tts')$ for some $({\tt c_1, \ldots, c_n})$ such that $\tc$ is not
one of the $\tc_i$, and $\tts\in \tuV$.
Roughly speaking, this says that $\tc$ is applicable at time $\tts$ if
has been declared 
not to have been revoked at $\ttt$ (under the assumption that 
at most one CRL is applicable at any given time).
Note that a certificate $\tc$ may be applicable at a time $\ttt$ outside
its validity interval.

We now define what it means for a 
formula 
$\phi$ to be true at a
run $r$ with respect to an interpretation $\pi = \la
\tlna,\tperm\ra$,
a key $\tk$, and a time $\con{t}$, written $\run,\int,\tk,\con{t}
\models \phi$, 
by induction on the structure of~$\phi$:
\bi
\item $\point \models  \tp \bdto \tq$ if $\intension{\tp}{L,\tk,\con{t}}
\supseteq \intension{\tq}{L,\tk,\con{t}}$, 

\item $\point \models \tc$  if $\tc\in \run(\con{t'})$ for some 
$\con{t}'\leq \con{t}$,

\item $\point \models \pperm(\tk_1,\tp,A)$ 
if for all $\tk_2\in \intension{\tp}{L,\tk_1,\tts}$ and all 
$\act\in \aint(A)$,
we have
$\tperm(\tk_1,\con{t},\tk_2,\act) \ge 1$,

\item $\point \models \pdel(\tk_1,\tp,A)$ if for all $\tk_2\in
\intension{\tp}{L,\tk_1,\tts}$ and all 
$\act\in \aint(A)$, 
we have $\tperm(\tk_1,\con{t},\tk_2,\act) = 2$,

\item $\point \models \now \in V$ if $\con{t} \in V$, 

\item $\point \models \valid(\tc)$ if $\tc$ is applicable at time $\tts$
in $r$,

\item $\point \models \phi \land \psi$ if $\point \models \phi$  and
$\point \models \psi$,

\item $\point \models \neg \phi$ if not $\point \models \phi$.

\ei

We write $r,\pi \sat \phi$ if $r, \pi, \tk, \ttt \sat \phi$ for all
principals $\tk \in K$ and all times $\ttt \in \Time$.

In the definitions so far, there is no connection between the run and
the interpretation. Intuitively, we would like the interpretation,
which contains information about the meaning of local names and
permissions and delegations, to be determined from the
information 
about the certificates that have been issued at each point in time
that is represented in the run. We connect these ideas by means of the
following definition. 
The interpretation $\int = \<L,P\>$ is {\em consistent\/} with a
run $\run$ if, 
for all times $\con{t}\in \nat$, \be
\item for all naming certificates $\tc ={\tt
(cert~\tk~\tn~\tp~V~ \la \tuR \ra)}$ in $\cup_{\tts'\leq \tts}
~\run(\tts')$, if $\tts\in V$ and $\tc$ is applicable at $\tts$ 
in $r$, 
then
$\intension{\tn}{L,\tk,\tts} \supseteq \intension{\tp}{L,\tk,\tts}$;
\item for all authorization certificates $\tc = {\tt
(cert~\tk~\tp}$ ${\tt 
\tuD~\tuA
~V~\la \tuR \ra)}$ in $\cup_{\tts'\leq
\tts}
~\run(\tts')$, if $\tts\in V$ and $\tc$ is applicable at $\tts$ 
in $r$, 
then
for all $\act \in \aint(\tuA)$ and all keys $\tk' \in
\intension{\tp}{L,\tk,\tts}$, we have \be
\item $\tperm(\tk,\con{t},\tk',\act) \ge 1$, 
\item if $\tuD= {\tt true}$ then
$\tperm(\tk,\con{t},\tk',\act) = 2$.
\ee
\ee

In general, an interpretation can be consistent with a run while
allowing facts to hold that do not follow from the certificates issued
in the run.  For an extreme example of this, let $\run$ be
the run in which no certificates are ever issued, and suppose that
$\pi$ 
is the {\em maximal interpretation},
where $\tlna(\tk,\tn,\con{t}) =
K$ and $\tperm(\tk,\con{t},\tk',\act) = 2$ 
for all keys $\tk$, $\tk'$, local names $\tn$, actions $\act$, and times
$\con{t}$.  Then $\int$ is consistent with $\run$. This is
undesirable. Intuitively, we would like facts concerning rights and
the meaning of local names to hold only if they are forced to do so by
some certificate. To enforce this, we restrict the interpretation to
be the minimal one consistent with the run.  Our technique for doing so
extends that used in \cite{HM01}.

Formally, define an order $\leq$ on 
$(K,N,\Act)$ interpretations 
by
$\la \tlna, \tperm, \ra \leq\la \tlna', \tperm' \ra$ if,
for all keys $\tk$, local names $\tn$, and times
$\con{t}$, we have $\tlna(\tk,\tn,\con{t})\subseteq
\tlna'(\tk,\tn,\con{t})$ and 
for all keys $\tk'$ and actions $\act$, we have 
$\tperm(\tk,\con{t},\tk',\act) \le \tperm'(\tk,\con{t},\tk',\act)$.
Thus, $\la L,P \ra \le \la L',P' \ra$ if, for all $\tn$,
$\tk$, and
$\tts$, at least as many keys are bound to $\tn$ by $\tk$ at time $\tts$
in $L'$ as in $L$.  In addition, at least as many keys are authorized to
perform  action $\act$ by $\tk$ at time $\tts$ in $P'$ as in $P$ and, of
those 
keys authorized to perform the action $\act$, at least as many can
delegate that authority in $P'$ as in $P$.

This order can
easily be seen to give the set of 
$(K,N,\Act)$ 
interpretations the structure of a
lattice.  Say that an element $\int$ of a set $S$ of interpretations
is {\em minimal in $S$} if $\int \leq \int'$ for all $\int'\in S$.

\pro\label{pro:minimal}
For every run $\run$, there exists a unique 
$(K,N,\Act)$ 
interpretation minimal
in the set of 
$(K,N,\Act)$ 
interpretations consistent with $r$.
\epro

\prf See the appendix.  \eprf

We write $\int_\run$ for the minimal interpretation consistent with
$\run$.
Intuitively, in the minimal interpretation consistent with $\run$, there
are no name bindings, permissions, or delegations that are not forced by
$\run$.  Enforcing the requirement that the interpretation should be the
minimal  one consistent with the run leads to a variant of the semantics
discussed above. We write $\run, \tk, \con{t} \models_c \phi$ if
$\run, \int_{\run}, \tk, \con{t} \models \phi$. We say that a
formula $\phi$ is {\em cl-valid} (with respect to $K, N, \Act$) in  
$\lspki(K,N,\Act)$, written $\sat_{\cl,K,N,\Act} \phi$, if $\run, \tk,
\con{t} \models_c \phi$ for all 
$(K,N,\Act)$-runs $\run$, keys $\tk \in K$, and times $\con{t}$.
(The ``cl'' stands for closed, since in \cite{HM01} the
semantics corresponding to $\sat_{\cl}$ is termed the {\em closed
semantics},
while the semantics corresponding to $\sat$ was termed the {\em open
semantics}.
We use ``cl'' here rather than ``c'' as in \cite{HM01} to denote the
closed semantics, to avoid confusion with $\tc$, which ranges over
certificates.) 
The closed semantics is the one of most interest to us here, since it 
enforces the desired close connection between the certificates actually
issued and the name bindings, permissions, and delegations.  The open
semantics is mainly used as a  stepping-stone to defining the closed
semantics. Validity with respect to the open semantics can also be
viewed as capturing what is guaranteed to hold, no matter what
additional certificates are issued.  Interestingly, as shown in
\cite{HM01}, validity with respect to the open and closed semantics
coincide for the logic LLNC.  We believe that they also coincide for the
logic $\lspki$, although we have not checked carefully.  

It is not hard to check that
the subscripts $N$ and $\Act$ play no role in $\sat_{\cl,K,N,\Ae}$; if a
formula is valid for some choice of $N$ and $\Act$ (for a fixed $K$),
then it is valid for all choices of $N$ and $\Act$.   However, as
observed in \cite{HM01} (where a complete axiomatization for LLNC was
provided), the axioms do depend on the choice of $K$; in particular,
they depend on the cardinality of $K$.  For example, suppose $K$ has
just one element, say $\tk$.  Then it is easy to see that 
$\sat_{\cl,K,N,\Act} \tn \bdto \tk \rimp \tk\s \tn \bdto \tk\s \tm$, for
all $\tn, \, \tm \in N$.    Since $\tn$ is bound
to $\tk$, it follows that 
the interpretation of $\tk\s \tn$ must be $\{\tk\}$, since
$\tk$ 
is the only key.  Thus, the interpretation of $\tk\s \tn$ must be a
superset of interpretation of $\tk\s \tm$, and so $\tk\s \tn \bdto \tk\s
\tm$ holds.  However, this argument depends critically on the assumption
that $K = \{\tk\}$.
The formula 
$\tn \bdto \tk \rimp \tk\s \tn \bdto \tk\s\tm$ 
is not valid if there are at least two keys: If $\tk'$ is a key
distinct from $\tk$, then it is possible that $\tk'$ is bound to $\tk\s
\tm$ and not bound to $\tk\s \tn$.  In light of this discussion,
we omit $N$ and $\Act$ from the subscript from here on in, but include
$K$ when it plays an important role.

We can now make precise the intuition that in the minimal interpretation
only name bindings and permissions and delegations forced by $r$ hold.
Define the {\em formula associated with the naming certificate\/}  
$\tc = {\tt (cert~\tk~\tn~\tp~V~\la \tuR \ra)}$  to be
$$\now \in V  \rimp (\tk\s \tn \bdto \tp).$$
Similarly, 
the {\em formula associated with the authorization
certificate\/}  $\tc = {\tt (cert~\tk~\tp~
\tuD~\tuA~
V~\la \tuR \ra)}$
is
\[\now \in V  \rimp 
[\pperm(\tk,\tp,\tuA) \land (\tuD \rimp \pdel(\tk,\tp,\tuA))].\]
Let $\form{\tc}$ be the formula associated with certificate $\tc \in
\Cert$.

\pro\label{pro:cert-form}
The interpretation $\pi$ is consistent with a run $r$ iff for all times
$\ttt \in \Time$, keys 
$\tk \in K$, and $\tc \in \Cert$, we have
$r, \pi, \tk, \ttt \sat \tc \land \valid(\tc) \rimp \form{\tc}$.
\epro
\prf Immediate from the definition of consistency.
\eprf

Note that it follows from Proposition~\ref{pro:cert-form} that if 
$\pi$ is an interpretation consistent with run $r$ and
a certificate $\tc \in \Cert$ was issued in $r$ at or before time $\ttt$
and
remains applicable at $\ttt$, then $r,\pi,\tk,\ttt \sat\phi_c$.
Moreover, the minimal interpretation associated with $r$ is the
minimal intepretation $\pi$ satisfying $r,\pi,\tk,\ttt \sat\phi_c$ for
all certificates $\tc \in \Cert$ that have been issued in $r$ at or
before time $\ttt$ and remain applicable at $\ttt$.  In this sense, 
the formulas associated with certificates precisely capture their
meaning.
It is also worth noting that 
$\sat_{\cl} \tc \land \valid(\tc) \rimp \form{\tc}$ for all $\tc \in \Cert$.

\commentout{
We will not attempt to provide a complete axiomatization of the logic
here, but note a few more valid formulas. It is not difficult to see
that all the formulas of LLNC are valid, and we refer the reader to
\cite{HM01} for a listing of those.\footnote{The validities of LLNC
depend on whether the set of keys is assumed to be finite or infinite,
and this dependency still applies here.} We focus here on a number of 
valid formulas that clarify the definitions or are useful
for the results to follow. 
\be 
\item $\now\in V_2 \land \now \in V_2 \liff \now \in V_1 \cap V_2$ 

\item $\valid(\tc)$, where $\tc$ is not revocable
\item 
$\tc \rimp (\now \in V \rimp \valid(\tc')$), where 
$\tc={\tt (crl~\tk~(
canceled
~\tc_1, \ldots,\tc_n))~V)}$
is a CRL, and $\tc'$ is a certificate 
revocable by $\tk$ but not equal to any of the $\tc_i$. 
\item $A_1 \supseteq A_1\cap A_2$ 

\item $\tp_1 \bdto \tp_2 \land A_1 \supseteq  A_2 \rimp 
(\pperm(\tk,\tp_1,A_1) \rimp \pperm(\tk,\tp_2,A_2))$

\item $\tp_1 \bdto \tp_2 \land A_1 \supseteq  A_2 \rimp 
(\pdel(\tk,\tp_1,A_1) \rimp \pdel(\tk,\tp_2,A_2))$
\ee 
}

\section{Soundness and completeness of tuple
reduction}\label{sec:verification}

We are now in a position to compare the SPKI tuple reduction rules to
our semantics.
Intuitively, we would like to understand 
tuple reduction
as drawing inferences about the run based on evidence
provided in the certificates presented to the verifier.
Thus, we want to show that all conclusions based on tuple
reduction are true under our semantics and, conversely, all 
valid conclusions about
bindings and authorizations that follow from the issuing of certain
certificates are derivable 
from those certificates
using tuple reductions.    
More precisely, we want to show that, given a finite set $C \union
C_R$ of certificates, where $C$ consists of 
naming and authorization certificates and $C_R$ consists of CRLs,
we have $\sat_{\cl} (\land_{\tc' \in C\union C_R} \tc') \rimp
\phi_{\tc}$ iff 
$\tau_{\tc}$ can be derived from the tuples in
$\Tuples{C,C_R}$, using the tuple reduction rules.
This can be broken up into two questions:
\begin{itemize}
\item {\em soundness}: if $\tau_{\tc}$ can be derived from the tuples in 
$\Tuples{C,C_R}$ using the tuple reduction rules, then 
$\sat_{\cl} (\land_{\tc' \in C\union C_R} \tc') \rimp
\phi_{\tc}$;
\item {\em completeness}: if $\sat_{\cl} (\land_{\tc' \in C\union C_R} \tc') \rimp
\phi_{\tc}$, then $\tau_{\tc}$ can be derived from the tuples in 
$\Tuples{C,C_R}$ using the tuple reduction rules.
\end{itemize}
Of course, whether soundness and completeness hold depend in large part
on {\em which\/} reduction rules are used.
We are particularly interested in questions such as whether R2$'$ and
R3$'$ are needed to derive all conclusions of interest about
certificates and, if not, what conclusions they can be used to derive.

Showing that the tuple reduction rules 
discussed earlier
are sound with respect to the
$\lspki$ semantics  is straightforward.  To make this precise, 
if $T$ is a set of tuples and $\tau$ is a tuple, we write
$T \deriv \tau$ if there exists a sequence of tuples
$\tau_1, \ldots , \tau_k$ such that $\tau_k = \tau$ and for each $i\leq
k$
either $\tau_i \in T$ or there exist $j,j'<i$ such that
$\tau_j + \tau_{j'} \longrightarrow \tau_i$ is an instance of
one of R1, R2, or R3.

\thm\label{thm:soundness}
Suppose that $C$ is a finite subset of $\Cert$, 
$C_R$ is a finite set of CRLs,
and
$\tc \in \Cert$.
If $\Tuples{C,C_R} \deriv \tau_{\tc}$, then  
$\models_c \left(\bigwedge_{\tc' \in C\cup C_R} ~\tc'\right) \rimp 
\phi_{\tc}$.
\ethm

Since Theorem~\ref{thm:soundness} is a special case of a more general
soundness
result, Theorem~\ref{thm:soundness1}, we defer the proof until after the
statement of the latter theorem.

Completeness is not at all straightforward; indeed, it
does not hold for 
R1--R3.
What does hold is a weak version of completeness.
To understand this in more detail, we need a few definitions.

A {\em concrete certificate\/} is  one whose
corresponding tuple has the form $\la \tk, \tn, \tk', [\ttt,\ttt] \ra$
(in the
case of naming certificates) or $\la \tk, \tk', \tuD, \{\act\},
[\ttt,\ttt] \ra$ 
(in the case of authorization certificates).  That is, concrete
certificates talk about the  keys that are bound to names and the 
keys that are authorized to perform certain actions, and are concerned
only with a single point in time and single actions.

The following result shows that R1, R2, and R3 suffice in a certain
sense to deal with concrete certificates.  
We say that a naming certificate  with 4-tuple
$\la \tk, \tn, \tp, \tuV \ra$ {\em subsumes\/} a 
naming certificate with 4-tuple
$\la \tk, \tn, \tp, \tuV' \ra$ if $ \tuV\supseteq \tuV'$.
Similarly an authorization certificate  with 5-tuple
$\la \tk, \tp, \tuD, \tuA, \tuV \ra$ 
subsumes 
an authorization certificate with 5-tuple
$\la \tk, \tp, \tuD', \tuA', \tuV' \ra$ if 
$\tuV\supseteq \tuV'$ and $\aint(\tuA) \supseteq \aint(\tuA')$ and $\tuD\rimp \tuD'$ 
is valid.  
(Recall that we have assumed
that it is possible to tell if one action expression is a superset of
another.)  Clearly if $\tc$ subsumes $\tc'$, then $\sat_{\cl} \phi_{\tc}
\rimp \phi_{\tc'}$.  

\thm\label{thm:concretecompleteness} If $\tc$ is a concrete certificate,
$C$ is a finite subset of $\Cert$, $C_R$ is a finite set of CRLs,
and $\models_c \left(\bigwedge_{\tc' \in C\cup C_R} ~\tc'\right) \rimp  
\phi_{\tc}$, then there exists a certificate $\tc'$ that subsumes $\tc$
such that $\Tuples{C,C_R} \deriv \tau_{\tc'}$.
\ethm

\prf See the appendix.  \eprf
 
Theorem~\ref{thm:concretecompleteness} tells us that if all we want to
do is to check if a given key is currently bound to a given name 
or if a given key is currently authorized to perform a given action, 
then R1--R3 essentially suffice.  
(Clarke et al.~\citeyear{CEEFMR} establish a closely related
result. They define the semantics of names operationally using just
R2' (which they call {\em rule composition}), and then establish that
all conclusions about whether a given key is bound to a given name can
be derived using just R2.)

However, R1--R3 do not suffice if we
want to do full-fledged reasoning about the consequences of a set of
certificates.  
For example, in general, R1--R3 may not suffice to draw a conclusion
of the form $\tk\s \tn \bdto \tp$, where $\tp$ is an arbitrary principal
expression, 
even though it may be a logical consequence of the
certificates issued.  
Such conclusions are of interest in the context of ``certificate
result certificates'' (\cite{spki1}, Section 6.7), which are new
certificates stating facts that are consequences of a set of
certificates previously issued. Certificate result certificates help
to reduce the amount of work a relying party needs to do when
processing a request.

There are three impediments to getting a full
completeness theorem that is a converse to Theorem~\ref{thm:soundness}.
The first two are easily dealt with by adding 
rules.

First, we want to get rid of the restriction that allows conclusions
only about keys.  R2 and R3 do not suffice for this.  For example, 
let $\tc_1$, $\tc_2$, and $\tc_3$ be 
irrevocable
naming certificates whose corresponding
4-tuples are 
$\<\tk_1, \tn, \tk_2\s \tm\s \tp, [\ttt,\ttt] \>$, $\<\tk_2, \tm, \tq,
[\ttt,\ttt] \>$, and 
$\<\tk_1, \tn, \tq\s \tp, [\ttt,\ttt] \>$, respectively.  
Clearly $\sat_{\cl} \tc_1 \land \tc_2 \rimp \phi_{\tc_3}$.
However, we cannot get $\tau_{\tc_3}$ from $\tau_{\tc_1}$ and
$\tau_{\tc_2}$, since the only rule that could possibly be of help, R2,
applies only to 4-tuples whose third argument is a  key.  To deal with
this problem, we need R2$'$.  Similarly, R3$'$ is needed to deal with
the analogous problem for R3.
We also need the following trivial axiom to deal with a special case:
\begin{itemize}
\item[R0.] $\longrightarrow \la \tk,\tn,\tk\s\tn,[0,\infty] \ra$.
\end{itemize}
(The fact that there is nothing to the left of the $\longrightarrow$ is
meant to indicate that the conclusion on the right-hand side can be
reached in all circumstances.)

A naming certificate is {\em point-valued\/} if its time interval has
the form
$[t,t]$.  A concrete certificate is point-valued  but a 
point-valued naming certificate can have as its third component an
arbitrary fully-qualified expression.
As we shall see, R0, R1, R2$'$, and R3$'$ essentially suffice to get us
conclusions about 
point-valued naming certificates, but do not suffice to get us
conclusions about 
arbitrary time intervals and sets of actions.  To see 
that they do not suffice, consider three irrevocable
naming certificates $\tc_1$, $\tc_2$, and $\tc_3$ whose
corresponding 4-tuples  are $\<\tk, \tn, \tp, [1,2] \>$, $\<\tk, \tn,
\tp, [3,4] \>$, and  $\<\tk, \tn, \tp, [1,4] \>$.
Clearly $\sat_{\cl} \tc_1 \land \tc_2 \rimp \phi_{\tc_3}$.
However, we cannot get $\tau_{\tc_3}$ from $\tau_{\tc_1}$ and
$\tau_{\tc_2}$, since none of the reduction rules increase the size of
intervals.  Similar issues arise with 5-tuples.

There are a number of ways to deal with this. Perhaps the simplest is
just to add the following reduction rule:
\be
\item[R4(a).] $\la \tk, \tn, \tp, \tuV_1 \ra + \la \tk, \tn, \tp, \tuV_2
\ra \longrightarrow \la \tk, \tn, \tp, \tuV_3 \ra$ if $\tuV_1 
\cup 
\tuV_2 \supseteq \tuV_3$.
\item[R4(b).] $\la \tk, \tp, \tuD, \tuA, \tuV_1 \ra + \la \tk, \tp,
\tuD, \tuA, \tuV_2 
\ra \longrightarrow \la \tk, \tp, \tuD, \tuA, \tuV_3 \ra$ if 
$\tuV_1 \cup \tuV_2 \supseteq \tuV_3$.
\item[R4(c).] $\la \tk, \tp, \tuD_1, \tuA_1, \tuV \ra + \la \tk, \tp,
\tuD_2, \tuA_2, \tuV 
\ra \longrightarrow \la \tk, \tp, \tuD_3, \tuA_3, \tuV \ra$ if 
$\tuD_3 \rimp \tuD_1 \land \tuD_2$ is a tautology and
$\aint(\tuA_1) \union \aint(\tuA_2) \supseteq \aint(\tuA_3)$.
\ee
Note that 
the rule 
\[\la \tk, \tn, \tp, \tuV_1 \ra \longrightarrow \la \tk, \tn, \tp, \tuV_3 \ra~~{\rm if}~~
\tuV_1  \supseteq \tuV_3\]
is a special case of
R4(a) (taking $\tuV_1 = \tuV_2$). Similar comments apply to R4(b) and R4(c).
How easy it is to apply R4(c) depends on how easy it is to check that
$\aint(\tuA_1) \union \aint(\tuA_2) \supseteq \aint(\tuA_3)$.
If action expressions can represent infinite sets of actions, this may
be nontrivial.  We do not address this issue here.
Of course, it is trivial to determine 
if $\tuV_1 \cup \tuV_2 \supseteq \tuV_3$.

We write $T \derivp \tau$ (\respc if $T \derivq \tau$) if there is a
derivation
of $\tau$ from $T$ using rules R0, R1, R2$'$, and R3$'$  (\respc R0, R1,
R2$'$, R3$'$, and R4).  There is no
difficulty showing that $\derivq$ is sound.
Note that the soundness of $\deriv$ and $\derivp$ follow as special
cases.

\thm\label{thm:soundness1}
Suppose that $C$ is a finite subset of $\Cert$, 
$C_R$ is a finite set of CRLs, and $\tc \in \Cert$.
If $\Tuples{C,C_R} \derivq \tau_{\tc}$, then  
$\models_c \left(\bigwedge_{\tc' \in C\cup C_R} ~\tc'\right) \rimp 
\phi_{\tc}$.  
In fact, if $\pi$ is an interpretation consistent with a run $r$ and
$\Tuples{C,C_R} \derivq \tau_{\tc}$, then  
$r,\pi \models \left(\bigwedge_{\tc' \in C\cup C_R} ~\tc'\right) \rimp 
\phi_{\tc}$.  
\ethm
Theorem~\ref{thm:soundness1} follows easily from the following two
propositions, 
whose straightforward proof we leave to the reader.
The first shows that the process of transforming a set of
certificates into a set of tuples corresponds to a valid inference in
the logic.

\pro 
Suppose that $\tc_1$ is a revocable certificate in $\Cert$, $\tc_2$ is a
CRL
such that $\tc_1$ is live with respect to $\tc_2$, and $\tc_3 = 
\tau(\tc_1,\tc_2)$.
Then if $\pi$ is an interpretation consistent with a run $r$ and
$\Tuples{C,C_R} \derivq \tau_{\tc}$, then  
$r,\pi \sat \tc_1 \land \tc_2 \rimp \form{\tc_3}$.  
Similarly, if $\tc$ is an irrevocable certificate in $\Cert$, then 
$r, \pi \sat \tc \rimp \form{\tc}$.  
\epro

The second proposition shows that single reductions 
are sound with respect to the closed semantics.

\pro\label{pro:redn-form}
Suppose that $\tc_1$, $\tc_2$, and $\tc_3$ are certificates and
$\tau_{\tc_1} + \tau_{\tc_2} \rightarrow \tau_{\tc_3}$ is an instance of
R1, R2$'$, R3$'$, or R4.  Then
$$\models_c \form{\tc_1} \land \form{\tc_2} \rimp \form{\tc_3}.$$ 
Moreover, if $\rightarrow \tau_{\tc}$ is an instance of R0, then
$\models_c \form{\tc} $.
\epro

As we show shortly,
R4 (together with R0, R1, R2$'$, R3$'$) suffices for completeness 
provided that the set of keys is infinite.  It does not suffice if the
set of keys is finite.  As we have seen, the set of valid formulas
depends on the 
cardinality of the
set $K$ of keys.  
Consider the very simplest case where $K$ consists of only one key
$\tk$.  Suppose that $\tm$ and $\tn$ are names in $N$.
Let $\tc$ and $\tc'$ be 
irrevocable
naming certificate whose 4-tuples  are 
$\la \tk, \tn,  \tk, \tuV \ra$ and 
$\la \tk, \tn, \tk\s \tm,\tuV\ra$, 
respectively.  It is easy to see that 
$\sat_{\cl,K} \tc \rimp \phi_{\tc'}$. This is true for essentially the
same reasons that $\sat_{\cl,K} \tn \bdto \tk \rimp \tk\s \tn \bdto \tk\s
\tm$. 
On the other hand,
$\sat_{\cl,K} \tc \rimp \phi_{\tc'}$ 
does not hold if $|K| \ge 2$ 
(and, in particular, if $K$ is infinite).
It easily follows that $\tau_{\tc'}$ is not derivable from $\{\tc\}$.
Additional derivation rules 
would be 
necessary to allow this derivation.

With a little more effort, examples like this can be given as long as
$K$ is finite.  That is, if $K$ is finite, then  
there exist finite sets $C \subseteq \Cert(K,N,\Ae)$ and $C_R \subseteq
\Cert_R(K,N,\Ae)$ and a naming certificate $\tc$ such that
$\sat_{\cl,K} (\land_{\tc' \in C \union C_R} \tc') \rimp \phi_{\tc}$, but
$\tau_{\tc}$ cannot be derived from $\Tuples{C,C_R}$ using R0, R1,
R2$'$,
R3$'$, R4.  
Nevertheless, these reduction rules are ``almost'' complete.
We show that, for any fixed set $C  \subseteq
\Certall(K,N,\Ae)$, as long as $|K| > |C| + |\tc|$, then for all sets
$C_R$, we have
$\sat_{\cl,K} (\land_{\tc' \in C \union C_R} \tc') \rimp \phi_{\tc}$
iff $\Tuples{C,C_R} \derivq \tc$.
(Here $|\tc|$ denotes the length of the certificate $\tc$ as a string of
symbols, and $|C|$ denotes the sum of the lengths of the certificates in
the finite set $C$.)  

\thm\label{thm:complete} If $\tc$ is a certificate,  
$C$ is a finite subset of $\Cert$, $C_R$ is a finite set of CRLs, 
$|K| > |C| + |\tc|$, and
$\models_{c,K} \left(\bigwedge_{\tc' \in C\cup C_R} ~\tc'\right) \rimp  
\phi_{\tc}$, then 
$\Tuples{C,C_R} \derivq 
\tau_{\tc}$;
moreover, if $\tc$ is a point-valued certificate, then $\Tuples{C,C_R}
\derivp \tau_{\tc'}$ 
for some certificate $\tc'$ subsuming $\tc$.
\ethm

\prf See the appendix. \eprf

Basically, Theorem~\ref{thm:complete} says that, by using R2$'$, R3$'$,
and R4 in the tuple reduction rules, we can derive all conclusions about
certificates, 
provided that $K$ is not ``small'', in the sense that $|K| \leq |C| +
|\tc|$.  This proviso is not at all unreasonable. The set $K$ is,
after all, intended to model the collection of potential public keys
being used by the principals. In order for such a set of keys to be
secure for signatures and encryption, it needs to be a very large set,
so as to render brute force attacks on encrypted messages impractical
(i.e., the key length needs to be large enough). By contrast, $C$
models a set of certificates generated by the principals and presented 
in a particular authorization request, and $\tc$ is a particular certificate. 
We do not expect the size of these to be of the same order of magnitude 
as $K$,  
so we would expect that $|K| \leq |C| + |\tc|$ will hold in practice. 
Another way to think about things is to understand $K$ as modelling the set of all keys
that could ever be used, if we allow the key length to be made 
arbitrarily large. In this case $K$ is an infinite set, so we immediately have that 
$|K| \leq |C| + |\tc|$. 
Thus, in a precise sense, the rules R0, R1, R2$'$, R3$'$, and R4
together give us all interesting conclusions about certificates.

\section{Related Work}\label{related}

As mentioned in the introduction, there have been a number of other recent approaches
to giving semantics to SPKI. In this section, we compare these approaches to ours. 

Howell and Kotz 
\citeyear{HK00a}
give a semantics to names 
that closely resembles the semantics of 
Abadi \citeyear{Abadi98}.  In particular, a name is associated
with a relation on possible worlds, rather than a set of keys as is the
case in our semantics.  We criticized Abadi's
semantics in our earlier paper \cite{HM01}; 
many of our 
criticisms 
apply to the
Howell-Kotz semantics as well. 
Perhaps most significantly, 
the semantics 
for name binding
(as Howell and Kotz themselves say) is rather opaque; it is hard to
explain 
exactly what the meaning of the relation on worlds is.  
Their logic also uses a ``speaks-for'' relation $\tp \rimp\tq$, where
$\tp,\tq$ are principal expressions, to capture both the binding
relationship between names (similar to our use of $\bdto$) and the
notion that $\tq$ has delegated certain rights to $\tp$. As we argued 
in \cite{HM01}, we believe that these are distinct notions that 
should be modeled using different constructs. Indeed, Howell and Kotz 
themselves note that their axiom 
\[ \tp \rimp \tq \supset \tp\s \tn \rimp  \tq\s \tn \] 
is ``suprisingly powerful'' and needs to be ``tempered'',
particularly when one considers the related notion ``speaks for on topic
$T$''.  
In our logic, the formula 
\[ \tp \bdto \tq \rimp \tp\s \tn \bdto  \tq\s \tn \] 
is valid, but this is not a problem for us, since we do 
not interpret ``$\bdto$'' as ``speaks-for''. 

Just as we do,
Howell and Kotz translate SPKI certificates to formulas in their logic.
There seem to be 
some 
problems with their translation, although
there are not enough formal details in the paper for us to be really
sure that these are in fact problematic.  
\begin{itemize}
\item To capture the difference between just being given permission to
perform an action and being given permission to delegate authority to
perform the action, they split a real 
principal $\tk$ into two principals
$\tk_u$ and $\tk_b$; $\tk_b$ is supposed to deal with the
situation where $\tk$ can delegate authority, while $\tk_u$ is
supposed to deal with the situation where $\tk$ cannot delegate
authority.  There is no formal semantics given to $\tk_u$ and $\tk_b$,
so it is not clear (to us) the extent to which this approach works.
\item Their approach to dealing with time seems to involve talking about
time explicitly in the formula.  But there is no type corresponding to
time in the semantics, so there is nothing to connect the statements
about time to the actual time.  Again, since the discussion of how time
is handled is quite informal (with no examples), it is hard to tell how
well it will work.
\end{itemize}
Howell and Kotz do 
claim
that their semantics is sound with respect to
tuple reduction; there is no discussion of completeness.

Aura \citeyear{Aura98} defines the notion of a {\em delegation network},
which is essentially a graphical description of a collection of
certificates.  Although he does not have a formal logic to reason about
authorization, given a delegation network $DN$, he does define 
a relation {\em authorizes}$_{DN}(\tk_1,\tk_2,\act)$,
which can be read as ``$\tk_1$ authorizes $\tk_2$ to perform $\act$''.
Given a collection of SPKI certificates, Aura defines a 
corresponding authorization network $DN$ and obtains a certain type of
soundness and 
completeness result for reduction.  Very roughly speaking, given a
collection of certificates with corresponding delegation network $DN$,
{\em authorizes}$_{DN}(\tk_1,\tk_2,\act)$ holds iff there is a sequence
of delegation networks $DN_1, \ldots, DN_m$ such that $DN_{i+1}$ is
obtained from $DN_i$ by certificate reduction in a precise sense and 
in $DN_m$ there is an explicit certificate issued by $\tk_1$ authorizing
$\tk_2$ to perform $\act$.  This is not a soundness and
completeness result in the logical sense that we have here, although it
provides
what can be viewed as an operational semantics for 
the SPKI 5-tuple reduction rule R1. 
One of the main differences from our work is that, where we deal with
principal expressions, 
Aura's framework deals with an explicitly given set of keys; there is
nothing
in his paper that corresponds to our discussion of name reduction. 
Aura also does not seem to have time or revocation in his
framework, nor does he consider the delegation bit.  
On the other hand, he does deal with threshold principals
(as do Howell and Kotz), while we do not.

Li \citeyear{Li00} has also considered the semantics of SPKI/SDSI. He
presents a logic program, and provides results showing that this
program derives the same set of concrete conclusions about SDSI names
as 4-tuple reduction and SDSI's name resolution procedure REF
\citeyear{RL96}. He also provides an extension of the logic program
intended to capture concrete conclusions about authorization
certificates.  However, he also does not explicitly treat timing and
revocation, as we have done, and does not consider general reasoning
such as that captured by our rules. On the other hand, he does discuss
threshold subjects, which we have not treated.  Another approach to
formally capturing SPKI's semantics is presented by Weeks
\citeyear{Weeks01} as an example of a more general framework for trust
management in a functional programming style. This work does not
attempt to prove any correspondence with the tuple reduction rules.

Also related to our results in this paper is the work of Clarke et 
al.~\cite{CEEFMR}, who consider the problem
of discovering ``certificate chains'', i.e. proofs that a given set of
certificates entails a given {\em concrete} certificate. Our notion of
derivation ``$\deriv$'' is similar to their notion of ``name-reduction
closure'', and Theorem~\ref{thm:concretecompleteness} is closely
related to a result they state (Theorem 1) concerning the completeness
of the name reduction closure. They do not explicitly take time and
revocation into account, as we have done, and they also do not
consider the more general sorts of consequences we have discussed. On
the other hand, they do characterize the computational complexity of
the certificate chain discovery problem for concrete certificates,
whereas we have not discussed the complexity of the inference problems
we treat.
We are currently examining the complexity of certificate
chain discovery in our more general setting.

\section{Discussion}\label{sec:discussion}
In this paper, 
we have given a semantics for SPKI certificates independent of 
the semantics given in terms of tuple reduction in the SPKI
documentation \cite{spki1,spki2}.  This allowed us to 
examine the extent to which the SPKI tuple reduction rules are
complete. 
The
SPKI documents are ambiguous as to the purpose of the tuple reduction
rules, and conflates their use for purposes of semantics, making
concrete authorization decisions, and general reasoning
(e.g., generating ``certificate result certificates'').  We have carefully
separated the three concerns here, 
and have
shown that the relations between them
are somewhat subtle. 
In particular, we have shown that extra reduction rules are needed in
order to do general reasoning about certificates.
Our main technical results show that, in a precise sense, the reduction
rules given in the SPKI document are complete with respect to concrete
certificates; adding a few more rules gives us an ``almost'' complete
system with respect to general reasoning about certificates.
This ``almost completeness'' result seems to be the best we can do
without having rules that take into account the cardinality of the set
of keys.

We need to be careful about the interpretation of the conclusions for
which we
have shown the tuple reduction rules to be complete.  The form of
conclusion associated with the tuple reduction rules
is $\sat_{\cl} (\land_{\tc' \in C \union C_R} \tc') \rimp \phi_{\tc}$,
where $\phi_{\tc}$ has the form $\now \in \tuV \rimp \phi_{\tc}'$ (with
the exact 
form of $\phi_{\tc}'$ depending on the type of certifcate that $\tc$ is).
This states
that 
$\phi_{\tc}'$ 
holds for all times in $\tuV$ at which additionally,
all
the certificates in $C \union C_R$ have been issued. 
Consider the certificate $\tc={\tt (cert~\tk~\tn~\tk'~[0,10])}$.
According to our semantics, if $\tc\in r(5)$, (i.e., this certificate
is issued at time 5), and no other certificates are issued in $r$, then
$\tk\s \tn \bdto \tk'$ holds in $r$ during the interval $[5,10]$, but
not during the interval $[0,4]$. 
On the other hand, the tuple reduction rules (trivially) allow the
derivation of the tuple $\la \tk,\tn, \tk',[0,10]\ra$, which suggests
that $\tk\s \tn \bdto \tk'$ also holds during the interval $[0,4]$.%
\footnote{We have followed the SPKI definitions closely, but we remark that 
we could modify the
definition of the tuple generated by a certificate and a CRL: if
$\tc$ has validity interval $[\ttt_1,\ttt_2]$, and the computation is done at $\ttt_0$, then we 
redefine the interval in $\tau(\tc,C_R)$ to be 
interval $[max(\ttt_0,\ttt_1),\ttt_2]$. All our completeness results would go through 
with this change, and it would result in a better match between 
our semantics and the tuple reduction rules.}  

Whether this difference matters depends on the use that is made of
derived tuples. For conclusions about the current time, the difference
does not matter, since agents will only be making authorization
decisions based on $\tc$ after the time at which $\tc$ is issued,
i.e., after time 5. However, for conclusions requiring reasoning about the past, the
difference may be an issue needing careful consideration. (Such reasoning occurs in proposals by 
Rivest \citeyear{Rivest98}, Stubblebine \citeyear{Stubblebine}, and
 Stubblebine and Wright \citeyear{SW96} that involve 
authorizing an action when the latest time for which the 
existence of a right to perform that action can be proved is ``sufficiently recent''.
Another example where reasoning about the past may
be important is where an auditor is verifying that an
authorization decision made in the past was justified by certificates
issued at the time of the decision.) 
As it does not appear that reasoning about the past was a 
significant concern in the design of SPKI, we 
do not pursue this further here.
Another contribution of this paper  is to show that nonmonotonic logic
is
not required in a logical modeling of revocation if one takes the
SPKI perspective that revocation is not a change of mind but a
revalidation. 
(This viewpoint is supported by the text in \cite[Section 5.2]{spki}:
``The CRL is ... a completion of the certificate, rather
than a change of mind.'')
The logic of this paper, like our earlier logic LLNC for
SDSI, is monotonic. 
This does not prevent some aspects of its semantics 
from
behaving nonmonotonically.
In particular, $L(\tk,\tn,\tts)$ may decrease as $\tts$ increases if,
for
example, a certificate is revoked at time $\tts' > \tts$ that was not
revoked at time $\tts$ or if the validity interval of a certificate
passes.  Similarly, the set of actions a principal is permitted to
perform may decrease over time.
Note also that 
the semantics does not require that if a certificate
appears on a CRL then it will also appear on all later CRLs issued
during the certificate's validity interval.%
\footnote{
Although it does not
appear to have been noted by the authors of SPKI, this allows CRLs to
be used to obtain temporary suspensions.}
All of these types of ``nonmonotonic'' behaviour are entirely consistent 
with the monotonic logic we have developed. 
We have focused on the SPKI reduction rules.  However, we
feel that the logic $\lspki$ will be useful for more general
reasoning about names and authorization in SPKI than just whether a
particular principal is authorized to perform some actions.  For
example, we may want to know which principals are authorized to perform
a
certain action, which actions a principal is allowed to perform, which
principals 
are bound to a particular name, and which names have a particular
principal
bound to them.   In \cite{HM01}, we showed how we could translate
queries about names (like the last two) into Logic Programming queries,
allowing us to take advantage of the well-developed Logic
Programming technology for answering such queries.  We believe that it
should be relatively straightforward to extend the translation so that
it can handle more general SPKI queries.  It also seems useful to try to
obtain a sound and complete axiomatization for the full logic $\lspki$.

We believe that it should not be difficult to get such a
complete axiomatization, using ideas from our earlier paper on SDSI, but
we have not pursued this question.
We have also not considered a number of features of SPKI, like threshold
subjects and 
the precise syntax of SPKI's tags (authorization expressions). 
It seems that our approach should be
extendible to handle these features, but we have not checked any details.

\appendix

\section{Appendix:Proofs}

\opro{pro:minimal}
For every run $\run$, there exists a unique 
$(K,N,\Act)$ 
interpretation minimal
in the set of 
$(K,N,\Act)$ 
interpretations consistent with $r$.
\eopro

\prf Let $\Ir$ consist of all interpretations consistent with
$r$.  $\Ir$ is nonempty, since the maximal interpretation is clearly
consistent with $r$.  Given an interpretation $\pi$, let $L_\pi$ denote
the $L$ component of $\pi$ and let $P_\pi$ denote the $P$ component of
$\pi$.  Define an interpretation $\pi_0$ by taking
$L_{\pi_0}(\tk,\tn,\ttt) = \inter_{\pi \in \Ir} L_\pi(\tk,\tn,\ttt)$ and
taking $P_{\pi_0}(\tk_1,\ttt,\tk_2,\act) = \min_{\pi \in \Ir}
P_\pi(\tk_1,\ttt,\tk_2,\act)$.   We now show that $\pi_0$ is the minimal
element of $\Ir$.

Clearly $\pi_0 \le \pi$ for all $\pi \in \Ir$.  Thus, we must only show
that $\pi_0 \in \Ir$.  First we must show that for all naming
certificates $\tc ={\tt (cert~\tk~\tn~\tp~V~ \la \tuR \ra)}$ in
$\cup_{\tts'\leq \tts} ~\run(\tts')$, if $\tts\in V$ and $\tc$ is
applicable at $\tts$ in $r$, then $\intension{\tn}{L_{\pi_0},\tk,\tts}
\supseteq \intension{\tp}{L_{\pi_0},\tk,\tts}$.  By definition
$\intension{\tn}{L_{\pi_0},\tk,\tts} = \inter_{\pi \in \Ir} 
\intension{\tn}{L_{\pi},\tk,\tts}$.  Moreover, since each $\pi \in \Ir$
is consistent with $r$, we have that $\intension{\tn}{L_{\pi},\tk,\tts}
\supseteq \intension{\tp}{L_{\pi},\tk,\tts}$ for $\pi \in \Ir$.
It follows immediately that $\intension{\tn}{L_{\pi_0},\tk,\tts} =
\inter_{\pi \in \Ir} \intension{\tn}{L_{\pi},\tk,\tts} \supseteq
\inter_{\pi
\in \Ir} \intension{\tp}{L_{\pi},\tk,\tts}$.  Thus, it suffices to show
that $\inter_{\pi \in \Ir} \intension{\tp}{L_{\pi},\tk,\tts} \supseteq
\intension{\tp}{L_{\pi_0},\tk,\tts}$.  This follows by an easy induction
on the structure of $\tp$.  For the base case, as we have already
observed, we have equality; for the general case the argument is almost
immediate from the definition.%
\footnote{We remark that this is a significantly simpler proof than that
given for the analogous minimality result for SDSI in our earlier paper
\cite[Theorem 3.1]{HM01}.}

Next we must show that $P_{\pi_0}$ is indeed a 
permission/delegation
assignment, that is, that if $\tperm_{\pi_0}(\tk_1,\con{t},\tk_2,\act) =
2$ and  $\tperm_{\pi_0}(\tk_2,\con{t},\tk_3,\act) = i$, then 
$\tperm_{\pi_0}(\tk_1,\con{t},\tk_3,\act) \ge i$.  But if
$\tperm_{\pi_0}(\tk_1,\con{t},\tk_2,\act) = 2$ and
$\tperm_{\pi_0}(\tk_2,\con{t},\tk_3,\act) = i$,  then 
$\tperm_{\pi}(\tk_1,\con{t},\tk_2,\act) = 2$ and
$\tperm_{\pi}(\tk_2,\con{t},\tk_3,\act) \ge i$ for all $\pi \in \Ir$.
Thus, $\tperm_{\pi}(\tk_1,\con{t},\tk_3,\act) \ge i$ for all $\pi \in
\Ir$, so $\tperm_{\pi_0}(\tk_1,\con{t},\tk_3,\act) \ge i$.

Finally, we must show that $P_{\pi_0}$ satisfies the second requirement
of consistency.  So suppose that $\tc = {\tt
(cert~\tk~\tp}$ ${\tt 
\tuD~\tuA
~V~\la \tuR \ra)}$ is an authorization
certificate in $\cup_{\tts'\leq \tts} ~\run(\tts')$, $\tts\in V$, and
$\tc$ is applicable at $\tts$ in $r$.  Then for all $\pi \in \Ir$,
for all $\act \in \aint(\tuA)$, and all keys $\tk' \in
\intension{\tp}{L_\pi,\tk,\tts}$, we have 
\be 
\item $\tperm_\pi(\tk,\con{t},\tk',\act) \ge 1$, 
\item if $\tuD= {\tt true}$ then
$\tperm_\pi(\tk,\con{t},\tk',\act) = 2$.
\ee
As we observed earlier, if $\tk' \in
\intension{\tp}{L_{\pi_0},\tk,\tts}$,
then $\tk' \in \intension{\tp}{L_{\pi},\tk,\tts}$ for all $\pi \in \Ir$.
Thus, $\tperm_\pi(\tk,\con{t},\tk',\act) \ge 1$ for $\pi \in \Ir$, so 
$\tperm_{\pi_0}(\tk,\con{t},\tk',\act) \ge 1$.  Moreover, if $\tuD= {\tt
true}$, then $\tperm_\pi(\tk,\con{t},\tk',\act) = 2$ for $\pi \in \Ir$,
so 
$\tperm_{\pi_0}(\tk,\con{t},\tk',\act) = 2$.  This completes the proof.
\eprf

\bigskip

We next prove the completeness results
(Theorems~\ref{thm:concretecompleteness} and~\ref{thm:complete}).  To do
this, we need some preliminary definitions and results.

Rules R2 and R3 are very similar in the way that they transform
principal expressions.  For the completeness proofs, it is convenient
to capture the commonalities by introducing a new type of tuple 
that we call a {\em 3-tuple}.  A 3-tuples has the  form $\la
\tp, \tq, \tuV\ra$ where $\tp,\tq$ are fully qualified principal
expressions and $\tuV$ is an interval.  Intuitively, this tuple says
that
$p$ is bound to $q$ during the time interval $\tuV$.

We have the following rules for reasoning about 3-tuples:
\be \item[R5.]  $\longrightarrow \la \tp, \tp,[0,\infty]\ra $ for all
fully
qualified 
principal expressions 
$\tp$.
\item[R6.] if $\la \tp, \tk_1\s \tn\s \tq,\tuV_1\ra + 
\la \tk_1,\tn,\tk_2,\tuV_2\ra \rightarrow
\la \tp, \tk_2\s \tq, \tuV_1\cap \tuV_2\ra $.
\ee

R5, like R0, is essentially an axiom.
R6 is somewhat in the spirit of R2, in that the third component
of the 4-tuple is a key, rather than an arbitrary full qualified name.
R6$'$ extends R6 in much the same way that R2$'$ extends R2.
\be
\item[R6$'$.] if $\la \tp, \tk_1\s \tn\s \tq,\tuV_1\ra + 
\la \tk_1,\tn,\tr,\tuV_2\ra \rightarrow
\la \tp, \tr\s \tq, \tuV_1\cap \tuV_2\ra $.
\ee

We abuse notation somewhat and continue to write $T \deriv \tau$ (\respc
$T \derivp \tau$) if $\tau$ can be derived from $T$ using the rules
R1--R3, R5, and R6 (\respc R0, R1, R2$'$, R3$'$, R5, R6$'$).
The following two propositions collect the key properties of 3-tuples.

\pro \label{prop:three-four}
Suppose that $i \in \{0,1\}$.
If $T$ is a set of 4- and
5-tuples such that $T\derivi \la \tp,\tq,\tuV_1\ra$, then 
\be 
\item[(a)] if $T\derivi \la \tk,\tn,\tp,\tuV_2\ra$ then $T\derivi \la
\tk,\tn,\tq,\tuV_1\cap \tuV_2\ra$;
\item[(b)] if $T\derivi \la \tk,\tp,
\tuD,
\tuA,\tuV_2\ra$ then $T\derivi \la
\tk,\tq,
\tuD, 
\tuA,\tuV_1\cap \tuV_2\ra$;
\item[(c)] if $T\derivi \la \tp',\tp,\tuV_2\ra$ then $T\derivi \la
\tp',\tq,\tuV_1\cap \tuV_2\ra$.  
\ee
\epro
\prf By a straightforward induction on the length of the derivation 
of $T\derivi \la \tp,\tq,\tuV_1\ra$.
\eprf

\pro \label{prop:three-ext} 
Suppose that $i \in \{0,1\}$.
For all fully qualified names $\tp,\tq$ and local names $\tn$, we have 
$T\derivi \la \tp,\tq,\tuV\ra$ iff $T\derivi \la \tp\s \tn,\tq\s
\tn,\tuV\ra$.
\epro 
\prf Again, by a straightforward induction on the length of the
derivation 
of $T\derivi \la \tp,\tq,\tuV_1\ra$.
\eprf

For convenience, we treat 3-tuples as certificates.  For the 3-tuple
$\tc = \la \tp, \tq, \tuV\ra$, we take $\tau_{\tc} = \tc$ and define
$\form{\tc}$ to be the formula $\now \in \tuV \rimp \tp \bdto \tq$.
This formula captures the intuition that the 3-tuple is essentially
saying that $\tp$ is bound to $\tq$.

The following theorem, whose proof is just like that of
Theorem~\ref{thm:soundness1}, says that $\derivp$ (and hence $\deriv$)
continues
to be sound in the presence of the rules for 3-tuples.

\thm \label{prop:sound:3} 
Suppose that $C$ is a finite subset of $\Cert$, $C_R$ is a finite set of
CRLs,
and $\tc$ is either a certificate in $\Cert$ or a 3-tuple.
If $\Tuples{C,C_R} \derivp \tau_{\tc}$, then  
$\models_c \left(\bigwedge_{\tc' \in C\cup C_R} ~\tc'\right) \rimp 
\phi_{\tc}.$
In fact, if $\pi$ is an interpretation consistent with a run $r$ and
$\Tuples{C,C_R} \derivp \tau_{\tc}$, then  
$r,\pi \models \left(\bigwedge_{\tc' \in C\cup C_R} ~\tc'\right) \rimp 
\phi_{\tc}$.  
\ethm

With this background, we are ready to prove
Theorem~\ref{thm:concretecompleteness}.

\othm{thm:concretecompleteness} If $\tc$ is a concrete certificate,  
$C$ is a finite subset of $\Cert$, $C_R$ is a finite set of CRLs,
and $\models_c \left(\bigwedge_{\tc' \in C\cup C_R} ~\tc'\right) \rimp  
\phi_{\tc}$, then there exists a certificate $\tc'$ that subsumes $\tc$
such that $\Tuples{C,C_R} \deriv \tau_{\tc'}$.
\eothm

\prf We prove the contrapositive.  Suppose that there does not exist a
certificate $\tc'$ that subsumes $\tc$ such that $\Tuples{C,C_R}
\deriv \tau_{\tc'}$.  We show that it is not the case that $\models_c
\left(\bigwedge_{\tc' \in C\cup C_R} ~\tc'\right) \rimp \phi_{\tc}$, by
showing that there is a run $r$ such that $r, \tk, \ttt \models_c 
\bigwedge_{\tc' \in C\cup C_R} ~\tc'$ for all times $\ttt$ but 
$r, \tk, \ttt_0 \not\models_c \phi_{\tc}$, where $\ttt_0$ is the time in
the concrete certificate $\tc$.

Construct $r$ as follows: define $r(0) = C\cup C_R$
and $r(n) = \emptyset$ for all $n>0$.   Clearly $r, \tk, \ttt \models_c
\bigwedge_{\tc' \in C\cup C_R} ~\tc'$ for all times $\ttt$.  To show
that $r, \tk, \ttt_0 \not\models_c \phi_{\tc}$, 
we need to identify the minimal interpretation consistent
with $r$. Consider the interpretation $\pi = \la L,P\ra$ 
defined as follows:
\be 
\item $L(\tk,\tn,\ttt)$ is the set of keys $\tk'$ such that 
$\Tuples{C,C_R} \deriv \la \tk, \tn, \tk',\tuV\ra$ where $\ttt \in
\tuV$.

\item $P(\tk, \ttt,\tk', \act) = 2$ if 
$\Tuples{C,C_R} \deriv \la \tk, \tk', \tuA, \true, \tuV\ra$ for some
action
expression $\tuA$ with $\act \in \aint(\tuA)$ and $\ttt \in \tuV$;
$P(\tk, \ttt,\tk', \act) = 1$ if 
$\Tuples{C,C_R} \deriv \la \tk, \tk', \tuA, \false, \tuV\ra$ for some 
$\tuA$ with $\act \in \aint(\tuA)$ and $\ttt \in \tuV$ and it
is not the case that $\Tuples{C,C_R} \deriv \la \tk, \tk', \tuA', \true,
\tuV'\ra$ for 
some $\tuA'$ with $\act \in \aint(\tuA')$ and $\ttt
\in \tuV'$; and
$P(\tk, \ttt,\tk', \act) = 0$ otherwise.
\ee

We must check that $P$ is a legimitate permission/delegation
assignment.  In particular, suppose that 
$\tperm(\tk_1,\con{t},\tk_2,\act) = 2$ and 
$\tperm(\tk_2,\con{t},\tk_3, \act) = i \ge 1$.  We must show that  
$\tperm(\tk_1,\con{t}, \tk_3, \act) = i$.   Since
$\tperm(\tk_1,\con{t},\tk_2,\act) =2$, $\Tuples{C,C_R} \deriv \la
\tk_1, \tk_2, \tuA_1, \true,\tuV_1\ra$ for some 
$\tuA_1$ with $\act \in \aint(\tuA_1)$ 
and some $\tuV_1$ containing $\ttt$. 
Similarly, we must have that $\Tuples{C,C_R} \deriv \la \tk_2, \tk_3,
\tuD_2,\tuA_2,
\tuV_2\ra$ where $\ttt \in \tuV_2$, for some 
$\tuA_2$ with $\act \in \aint(\tuA_2)$, some $\tuD_2$ such
that if $i = 2$ then $\tuD=\true$, and some $\tuV_2$ containing $\ttt$.
Using R1, it follows that $\Tuples{C,C_R} \deriv \la \tk_1, \tk_3,
\tuD_2,\tuA_1 \inter \tuA_2, 
\tuV_1 \inter \tuV_2\ra$.  Since $\act \in \aint(\tuA_1 \inter \tuA_2)$ and 
$\ttt \in \tuV_1 \inter \tuV_2$, it follows that 
$\tperm(\tk_1,\con{t},\tk_3,\act) \geq i$, as desired.

Next we must show that this interpretation is the minimal one consistent
with $r$.  We first establish that it is consistent.
Thus, we must show  that the formulas associated with naming
and authorization certificates are satisfied in $r$.
To do this, we use the following lemma.  (This lemma was our
main motivation for introducing 3-tuples.) 

\lem\label{3red}  If $\tp$ is a  fully qualified principal expression,
then $\tk'\in \intension{\tp}{L,\tk,\ttt}$ iff \\ 
$\Tuples{C,C_R} \deriv \la \tp, \tk',V\ra$ for some interval $\tuV$
containing $\ttt$. 
\elem

\prf We proceed by induction on the structure of $\tp$. If
$\tp$ is a key $\tk'$ then, by definition,
$\intension{\tk'}{L,\tk,\ttt} = \{\tk'\}$.
By R5,  $\Tuples{C,C_R}\deriv \la \tk',\tk', [0,\infty]\ra$.
For the converse, note that a straightforward  induction on the
length of the derivation shows that 
if $\Tuples{C,C_R}\deriv \la
\tk',\tq, \tuV\ra$, then $\tq = \tk'$ and $\tuV = [0,\infty]$. 

For the inductive step, suppose that $\tp = \tq\s \tn$.  We 
first suppose that $\tk'\in \intension{\tp}{\tk,L,\ttt}$, and show that
$\Tuples{C,C_R} \deriv \la \tp, \tk',\tuV\ra$ for some interval $\tuV$
containing $\ttt$. By 
definition of $\intension{\tp}{L,\tk,\ttt}$, there exists
$\tk_1 \in \intension{\tq}{L,\tk,\ttt}$ such that $\tk' \in
L(\tk_1,\tn,\ttt)$. By the inductive hypothesis, we have
$\Tuples{C,C_R}\deriv \la \tq, \tk_1,\tuV_1\ra$ for some interval
$\tuV_1$ containing $\ttt$. 
By Proposition~\ref{prop:three-ext}, it follows that
$\Tuples{C,C_R}\deriv \la \tq\s \tn , \tk_1\s \tn ,\tuV_1\ra$.  By
definition of $L$ and the fact that $\tk' \in L(\tk_1,\tn,\ttt)$, we
have $\Tuples{C,C_R}\deriv \la \tk_1, \tn, \tk' ,\tuV_2\ra $ for some
$\tuV_2$ containing $\ttt$.  By R6, it follows that
$\Tuples{C,C_R}\deriv \la \tq\s \tn, \tk', \tuV_1\cap \tuV_2\ra$.
This is what we need, since $\ttt\in \tuV_1\cap \tuV_2$.

For the converse, suppose that $\Tuples{C,C_R} \deriv \la \tp,
\tk',\tuV\ra$
for some interval $\tuV$ containing $\ttt$.  We must show that $\tk'\in
\intension{\tp}{L,\tk,\ttt}$. Since $\Tuples{C,C_R} \deriv \la \tp,
\tk',\tuV\ra$ and $\tp = \tq\s \tn$, it follows that there must be some
key $\tk_1$ such that $\Tuples{C,C_R} \deriv \la \tp, \tk_1\s \tn,
\tuV_1\ra$ and $\Tuples{C,C_R} \deriv \la \tk_1, \tn, \tk',\tuV_2\ra$,
where $\tuV = \tuV_1 \inter \tuV_2$.  By the definition of $L$, we have
that  $\tk' \in L(\tk_1,\tn, \ttt)$.  Since
$\Tuples{C,C_R} \deriv \la \tq\s \tn, \tk_1\s \tn, \tuV_{m-1}\ra$, by
Proposition~\ref{prop:three-ext}, we also have $\Tuples{C,C_R} \deriv
\la \tq, \tk_1,\tuV_{m-1}\ra$.  By the induction
hypothesis, it follows  that $\tk_1\in \intension{\tq}{L,\tk,\ttt}$.  It
is now
immediate that $\tk'\in \intension{\tp}{L,\tk,\ttt}$. 
\eprf

Continuing with the proof of the theorem, recall that we must show that
the formulas associated with certificates issued in $r$ are satisfied in
$r$.  Suppose that 
$\tc ={\tt (cert~\tk~\tn~\tp~V~\la \tuR\ra)}$ 
is a naming
certificate in $\cup_{\tts'\leq \tts} ~\run(\tts')$ that is
applicable 
in 
$r$ at time $\ttt\in \tuV$.  We need to show that $\intension{\tn}{L,
\tk,\ttt} \supseteq \intension{\tp}{L, \tk,\ttt}$.  For this, note
that either $\tk_r$ is not present in $\tc$ and and $\tau(\tc) = \la
\tk, \tn, \tp, \tuV\ra$, or there exists a CRL $\tc_R\in C_R$ with
respect to which $\tc$ is live and $\tau(\tc,\tc_R) = \la \tk, \tn,
\tq, \tuV'\ra$, where again $\ttt\in \tuV'$.  In either case, it follows
that
$\Tuples{C,C_R} \deriv \la \tk, \tn, \tp, \tuV_1\ra$ for some $\tuV_1$
containing $\ttt$.  

We now want to show that if $\Tuples{C,C_R} \deriv \la \tk, \tn, \tp,
\tuV_1\ra$, then $\intension{\tn}{L, \tk,\ttt} \supseteq
\intension{\tp}{L, \tk,\ttt}$. 
Suppose that $\tk'\in \intension{\tp}{L,\tk,\ttt}$.  By
Lemma~\ref{3red}, we have that 
$\Tuples{C,C_R} \deriv \la \tp, \tk',\tuV_2\ra$ for some $\tuV_2$
containing
$\ttt$.  Since $\Tuples{C,C_R} \deriv \la \tk, \tn, \tp,
\tuV_1\ra$, by Proposition~\ref{prop:three-four}, it follows that 
$\Tuples{C,C_R} \deriv \la \tk,\tn, \tk',\tuV_1\cap \tuV_2\ra$.  Using
the
definition of $L$, it follows that $\tk' \in
\intension{\tn}{L,\tk,\ttt}$, as desired.

Next, suppose that  
$\tc ={\tt (cert~\tk~\tp~\tuD~\tuA~\tuV~\la \tuR\ra)}$ 
is
an authorization certificate in
$\cup_{\ttt'\leq \ttt}r(\ttt')$ that is valid
in $r$ at time $\ttt\in \tuV$.   As in the case of naming certificates,
it follows that there is some interval $\tuV_1$ such that $\ttt \in
\tuV_1$
and $\Tuples{C,C_R} \deriv \la \tk, \tp, \tuD, \tuA, \tuV_1\ra$.  
Suppose that $\tk'\in \intension{\tp}{L,\tk,\ttt}$. By Lemma~\ref{3red},
$\Tuples{C,C_R} \deriv \la \tp, \tk',\tuV_2\ra$ for some $\tuV_2$
containing $\ttt$.  By Proposition~\ref{prop:three-four},
$\Tuples{C,C_R} \deriv \la \tk,\tk', \tuD,\tuA,\tuV_1\cap \tuV_2 \ra$. From 
the definition of $P$, it follows that 
\[ r,\pi, \tk,\ttt\models \pperm(\tk,\tp,\tuA) \land (\tuD\rimp
\pdel(\tk,\tp,\tuA)), \]  
as required. 

To see that $\la L, P \ra$ is in fact the minimal interpretation
consistent with $r$, suppose that $\pi' = \la
L',P'\ra$ is another interpretation consistent with $r$. 
We need to show that $\pi \leq \pi'$. We first show that 
$L\leq L'$. Suppose that $\tk'\in L(\tk,\tn,\ttt)$. 
By definition of $L$, $\Tuples{C,C_R} \deriv 
\la \tk, \tn, \tk',\tuV\ra$  
for some interval $\tuV$ containing $\ttt$. 
Since $\pi'$ is consistent with $r$, by
Theorem~\ref{prop:sound:3},  
\[ r, \pi' \models \left(
\bigwedge_{\tc' \in C\cup C_R} \tc'\right)
\rimp
(\now \in \tuV \rimp \tk\s \tn \bdto \tk').\] 
Moreover, we 
have that
$r, \pi', \tk, \ttt \models 
\bigwedge_{\tc' \in C\cup C_R} \tc'$, by defintion of $r$.
Since $\ttt \in \tuV$, it follows that  $ r, \pi',\tk, \ttt
\models \tk\s \tn \bdto \tk'$, so $\tk' \in L'(\tk,\tn,\ttt)$.
Thus, $L \le L'$.  The argument that $P \le P'$ is very similar, and
is left to the reader. 

It remains to show that $r, \tk, \ttt_0 \not \models_c
\phi_{\tc}$.  To see this, suppose first that $\tc$ is a naming
certificate such that  $\tau_{\tc} = 
\la \tk,\tn,\tk',[\ttt_0, \ttt_0]\ra$. Since
$\tc$ is not subsumed by any 4-tuple derivable from $\Tuples{C,C_R}$,
it is immediate from the definition of $L$ that $\tk' \not \in
L(\tk,\tn,\ttt_0)$.  
{F}rom this it follows that $\point_0 \not \models_c
\phi_{\tc}$.  The argument for authorization certificates is
similar. 
\eprf 

We now prove Theorem~\ref{thm:complete}.

\othm{thm:complete} If $\tc$ is a certificate,  
$C$ is a finite subset of $\Cert$, $C_R$ is a finite set of CRLs, and
$|K| > |C| + |\tc|$, and
$\models_{c,K} \left(\bigwedge_{\tc' \in C\cup C_R} ~\tc'\right) \rimp  
\phi_{\tc}$, then 
$\Tuples{C,C_R} \derivq 
\tau_{\tc}$;
moreover, if $\tc$ is a point-valued certificate, then $\Tuples{C,C_R} \derivp \tau_{\tc'}$
for some certificate $\tc'$ subsuming $\tc$.
\eothm

\prf 
The proof proceeds much in the same spirit as the proof of
Theorem~\ref{thm:concretecompleteness}, using 3-tuples. 
Suppose that it is not the case that
$\Tuples{C,C_R} \derivq \tau_{\tc}$ (or it is not the case that
$\Tuples{C,C_R} \derivp \tau_{\tc}$, in the case that $\tc$ is a
point-valued naming certificate).   
Again the idea is to construct a run $r$ such that $r, \tk, \ttt
\models_c  
\bigwedge_{\tc' \in C\cup C_R} ~\tc'$ for all times $\ttt$ and 
that $r, \tk, \ttt_0 \not\models_c \phi_{\tc}$ for some time $\ttt_0$.
The construction of $r$ is somewhat more complicated than in the case of
Theorem~\ref{thm:concretecompleteness}.

Given a principal $\tp$, let $\Cl(\tp)$ 
be the smallest set 
$S'$
of principal expressions containing $\tp$,
such that if $\tp\s \tn \in S'$ then $\tp\in S'$.
It is
easy to see 
that $|\Cl(\tp)| \le |\tp|$, where $|\tp|$ is the length of $\tp$ viewed
as a string of symbols.  If $C'$ is a set of certificates, let $\Cl(C')$
be the union of $\Cl(\tp)$, for all the principal expressions $\tp$ that
appear in a certificate in $C'$
as well as the principal expressions $\tk\s \tn$ for 
4-tuples $\la \tk,\tn,\tq,V\ra$ in $C'$. 
Again,  it should be clear that
$|\Cl(C')| \le |C'|$.  We will be interested in the set $S = \Cl(C
\union \{\tc\})$ of pricipal expressions.  Note that
since we have assumed that $|K| > |C| + |c|$, it follows that $|K| >
|S|$.

Let $T$ be the set of time points containing $0$, $\infty$, and the
left and right components of each interval in $C\cup C_R$.  Note first
that only a finite number of intervals $\tuV$ can appear in a tuple $\la
\tp_0,\tq_0,\tuV\ra$ 
generated from $\Tuples{C,C_R}$, since every interval in a tuple
generated has both left and right components in $T$. 
Hence, $T$ is finite.
Let $\V$ be the
set of all point intervals $[\ttt,\ttt]$ where $\ttt\in T - 
\{\infty\}$, together with all intervals $[\ttt_1+1,\ttt_2-1]$ such that
$\ttt_1, \ttt_2 \in T$, $\ttt_1<\ttt_2$ and for no $\ttt\in T$ do we
have $\ttt_1<\ttt<\ttt_2$. (We take $\infty -1=\infty$).  Note that
the intervals in $\V$ are pairwise disjoint and 
span $\Time$. 
Moreover,
if $\tuV$ is the validity interval of a 
tuple derivable from $\Tuples{C,C_R}$ and $\tuW\in \V$, then either
$\tuV\cap \tuW = \emptyset$ or $\tuV\supseteq \tuW$.  
Finally, note that $\V$ is finite, since $T$ is.
For each interval 
$\tuW\in \V$, define the equivalence relation $\approx_{\tuW}$ on $S$ by
$\tp_1 \approx_{\tuW} \tp_2$ if both $\Tuples{C,C_R} \derivp
\la \tp_1,\tp_2,\tuV'\ra$ for some $\tuV'\supseteq \tuW$ and
$\Tuples{C,C_R} \derivp 
\la \tp_2,\tp_1,\tuV''\ra$ for some $\tuV''\supseteq \tuW$. We write
$[\tp]_{\tuW}$ for the 
equivalence class of $\approx_{\tuW}$ containing $\tp$.

Let $X$ be a set of keys in $K$ of cardinality  $|S|$.  For each
interval $\tuW\in \V$ and equivalence class $x$ of $\approx_{\tuW}$,
choose a key $\tk_{x,\tuW}$ in $X$ in such a way that if $x \ne y$, then
$\tk_{x,\tuW} \ne \tk_{y,\tuW}$.  
Without loss of generality, we may assume that 
$\tk_{[\tk]_{\tuW},\tuW} = \tk$ for each key $\tk \in S$.  (Note that
$[\tk]_{\tuW} = \{\tk\}$ since, as we observed earlier, if
$\Tuples{C,C_R} \derivp \la \tk,\tq,\tuV\ra$, then we must have $\tp = \tk$.) 
For ease of exposition, we write $\tk_{\tp,\tuW}$ rather than
$\tk_{[\tp]_{\tuW},\tuW}$.

Consider the run $r$ where no certificates are issued after time 0 and
$r(0)$ consists of the following certificates:
\be 
\item all certificates in $C \cup C_R$, 
\item for each $\tuW \in \V$, principal expression 
of the form $\tp\s \tn$ in $S$, 
principal expression $\tq\in S$ such that 
$\Tuples{C,C_R} \derivp \la \tp\s \tn,\tq,\tuV' \ra$ for some 
$\tuV'\supseteq \tuW$, 
the naming certificate 
${\tt (cert~\tk_{\tp,\tuW}~\tn~\tk_{\tq,\tuW}~\tuW)}$,
\item for each $\tuW \in \V$, principal expression $\tq \in S$, and key
$\tk \in S$ such that $\Tuples{C,C_R} \derivp \la \tk, \tq,\tuD, \tuA,
\tuV'
\ra$ for some  
$\tuV'\supseteq \tuW$, 
the authorization certificate ${\tt (cert~\tk~
\tk_{\tq,\tuW}~ \tuD~ \tuA ~\tuW)}$.
\ee
Note that this is a finite set of certificates since $\W$ and $S$ are a finite sets.

We now construct an interpretation $\pi = \la L,P \ra$ consistent
with $r$. 
For $\ttt\in \nat$, let $\tuWt$ be the unique interval in $\V$
containing $\ttt$.  Define $L(\tk,\tn,\ttt) = \emptyset$ unless $\tk$
is
$\tk_{\tp,\tuWt}$ for some principal
expression $\tp$; $L(\tk_{\tp,\tuWt}, \tn, \ttt) =
\{\tk_{\tq,\tuWt}: 
\Tuples{C,C_R} \derivp \la \tp\s \tn,\tq,\tuV'\ra, \tuV'\supseteq
\tuWt\}$.  $L(\tk_{\tp,\tuWt}, \tn, \ttt)$ is well defined
since, if $\tp'\approx_{\tuWt}\tp$, then by
Proposition~\ref{prop:three-ext}, 
$(\tp')\s \tn \approx_{\tuWt}\tp\s \tn$, hence
$\Tuples{C,C_R} \derivp \la \tp\s \tn,\tq,\tuV'\ra$ 
iff  $\Tuples{C,C_R} \derivp \la (\tp')\s \tn,\tq,\tuV'\ra$.
Similarly, $P(\tk,\ttt,\tk',\act) = 0$ unless $\tk'$ 
is
$\tk_{\tq,\tuWt}$, for some principal expression $\tq$.   
If $\Tuples{C,C_R} \deriv \la \tk, 
\tq,\true,\tuA,\tuV\ra$, where 
$\act\in \aint(\tuA)$ and $\tuV \supseteq \tuWt$, then 
$P(\tk,\ttt,\tk_{\tq,\tuWt}, \act) = 2$; 
if $\Tuples{C,C_R} \deriv \la \tk, \tq,\false,\tuA,\tuV\ra$, where 
$\act\in \aint(\tuA)$ and $\tuV \supseteq \tuWt$ and it is not the case
that
$\Tuples{C,C_R} \deriv \la \tk,
\tq,\true,\tuA',\tuV'\ra$ for some $\tuA'$ such that $\act \in \aint(\tuA')$
and $\tuV' \supseteq \tuWt$, then
$P(\tk,\ttt,\tk_{\tq,\tuWt}, \act) = 1$;
otherwise, $P(\tk,\ttt,\tk_{\tq,\tuWt}, \act) = 0$.  

We want to show that $\pi$ is the minimal interpretation consistent with
$r$.  We first need an analogue of Lemma~\ref{3red}.

\lem\label{3redp} For all~$\tp\in S$, keys $\tk$, and times $\ttt$,
$\intension{\tp}{L,\tk, \ttt} = \{
\tk_{\tq,\tuWt}: \tq \in S, \, \Tuples{C,C_R} \derivp
\la \tp,\tq,\tuV'\ra, \, \tuV'\supseteq \tuWt\}$. 
\elem
\prf We proceed by induction on
the construction of $\tp$.  If $\tp$ is a key or an expression
of the form $\tk\s \tn$, the claim follows trivially from the
definitions.  
Note we cannot have $\tp = \tn$ for a local name $\tn$, by the
construction of $S$.
So suppose that $\tp$ has the  form $\tq\s \tn$. 
We first show that if $\tk'\in \intension{\tq\s \tn}{L,\tk,
\ttt}$, then there exist $\tr \in S$ and $\tuV' \supseteq \tuWt$ such
that
$\tk'=\tk_{\tr,\tuWt}$ and
 $\Tuples{C,C_R} \derivp \la \tq\s \tn,\tr,\tuV'\ra$.
By definition of $\intension{\tq\s \tn}{L,\tk,
\ttt}$, we have that $\tk'\in L(\tk'',\tn,\ttt)$ for some key $\tk'' \in
\intension{\tq}{L,\tk, \ttt}$.  By the induction hypothesis, there
exists $\tq_1\in S$ such that $\tk'' = \tk_{\tq_1,\tuWt}$ and 
$\Tuples{C,C_R} \derivp \la \tq,\tq_1,\tuV' \ra$ for some
$\tuV'\supseteq \tuWt$. It follows from Proposition~\ref{prop:three-ext}
that
$\Tuples{C,C_R} \derivp \la \tq\s \tn,\tq_1\s \tn,\tuV' \ra$. 
By definition of $L$, since $\tk'\in
L(\tk_{\tq_1,\tuWt},\tn,\ttt)$,  
there exists a principal expression $\tq_2$ such that 
$\Tuples{C,C_R} \derivp \la \tq_1\s \tn,\tq_2,\tuV''\ra$, where $\tuV''
\supseteq \tuWt$ and $\tk' = \tk_{\tq_2,\tuWt}$. 
By Proposition~\ref{prop:three-four}, it follows that 
$\Tuples{C,C_R} \derivp \la \tq\s \tn,\tq_2,\tuV'\cap \tuV'' \ra$. 
Since $\tuV'\cap \tuV'' \supseteq \tuWt$, we are done. 

For the converse, suppose that $\Tuples{C,C_R} \derivp \la \tq\s
\tn,\tr,\tuV' \ra$, where $\tr \in S$ and $\tuV'\supseteq \tuWt$. Since 
$\Tuples{C,C_R} \derivp \la \tq,\tq,[0,\infty] \ra$, we have that
$\tk_{\tq,\tuWt} \in \intension{\tq}{L,\tk,
\ttt}$, by the induction hypothesis. Moreover, by definition of $L$, we
have 
$\tk_{\tr,\tuWt}\in L(\tk_{\tq,\tuWt},\tn)$. 
It follows that $\tk_{\tr,\tuWt}\in\intension{\tq\s \tn}{L,\tk,
\ttt}$. \eprf

We use Lemma~\ref{3redp} to show that $\pi$ is consistent with $r$. 
Consider a naming certificate $\tc = {\tt
  (cert~\tk_1~\tn~\tp~\tuV_1~\la \tk_r \ra)} \in C$ that is applicable at time $\ttt\in \tuV_1$.  
Either $\tc$ is irrevoccable, or there exists a 
CRL $\tc'$ in $r(0)$ with validity interval $\tuV_2$ such that $\ttt\in
\tuV_2$ and $\tc$ is live with respect to $\tc'$.
As in the proof of Theorem~\ref{thm:concretecompleteness}, 
in either case, there is an
interval $\tuV' \subseteq \tuV_1$ containing $\ttt$ such that 
$ \Tuples{C,C_R} \deriv \la \tk_1,\tn,\tp, \tuV'\ra $.  Since $\ttt \in
\tuV_1$, we must have $\tuV'  \supseteq \tuWt$.
By R5, $\Tuples{C,C_R}
\derivp \la \tk_1\s \tn, \tk_1\s \tn, [0,\infty]\ra$ so, using R6$'$, it
follows that $\Tuples{C,C_R}\derivp \la\tk_1\s \tn, \tp,\tuV'\ra$.  Now
if $\tk_2\in \intension{\tp}{L,\tk_1, \ttt}$, then by Lemma~\ref{3redp},
$\tk_2 = \tk_{\tq,\tuWt}$ for some $\tq \in S$ such that
$\Tuples{C,C_R} \derivp \la\tp,\tq,\tuV''\ra$, where $\tuV'' \supseteq
\tuWt$. 
It follows from Proposition~\ref{prop:three-four}(c) 
that $\Tuples{C,C_R}\derivp \la \tk_1\s \tn,\tq,\tuV'\cap \tuV'' \ra$.
Since $\tuV' \inter \tuV'' \supseteq \tuWt$, 
it follows from the defintion of $L$ that 
$\tk_2=\tk_{\tq,\tuWt}\in \intension{\tn}{L,\tk_1, \ttt}$, as required. 

Next, consider certificates of the form ${\tt
 (cert~\tk_{\tp,\tuW}~\tn~\tk_{\tq,\tuW}~\tuW)}$, where
$\tuW \in \V$,  $\tp\s \tn, \tq \in S$,
and $\Tuples{C,C_R} \derivp \la \tp\s \tn,\tq,\tuV' \ra$ for some
$\tuV'\supseteq \tuW$.  Since such certificates are irrevocable, they
are always 
applicable.
If $\ttt\in W$, we have
$\intension{\tn}{L,\tk_{\tp,\tuW}, 
\ttt}=L(\tk_{\tp,\tuW},\tn,\ttt) \supseteq \{ \tk_{\tq,\tuW}\}
=\intension{\tk_{\tq,\tuW}}{L,\tk_{\tp,\tuW}, \ttt}$ by
construction. 

Next, suppose that  
        $\tc ={\tt (cert~\tk~\tp~\tuD~\tuA~\tuV~\la \tuR\ra)}$
is an authorization certificate in $C$ that
is applicable 
in $r$ at time $\ttt\in \tuV$.  As in the case of naming
certificates,
it follow that there is some interval $\tuV_1$ such that $\ttt \in
\tuV_1$ and $\Tuples{C,C_R} \derivp \la \tk, \tp, \tuD, \tuA,
\tuV_1\ra$.   Suppose that $\tk'\in \intension{\tp}{L,\tk,\ttt}$. By
Lemma~\ref{3redp},  $\tk' = \tk_{\tq,\tuWt}$ for some $\tq \in
S$ such that $\Tuples{C,C_R} \derivp \la \tp,\tq,\tuV_2\ra$ for some 
$\tuV_2 \supseteq \tuWt$.  By Proposition~\ref{prop:three-four}, it
follows that 
$\Tuples{C,C_R} 
\derivp
\la \tk,\tq, \tuD,\tuA,\tuV_1\cap \tuV_2 \ra$. From 
the definition of $P$
and the fact that $\tk' = \tk_{\tq,\tuWt}$,
it follows that 
\[ r,\pi, \tk,\ttt\models 
\pperm(\tk,\tk',\tuA) \land (\tuD\rimp \pdel(\tk,\tk',\tuA)), \]  
as required. 

Finally, consider an authorization certificate in $r(0)$ of the form 
 ${\tt (cert~\tk~\tk_{\tq,\tuW}~ \tuD~ \tuA ~\tuW)}$.  
By definition, $\Tuples{C,C_R} \derivp \la \tk, \tq,\tuD, \tuA, \tuV'
\ra$ for some $\tuV'\supseteq \tuW$.  Again, it is immediate from the
 definition of $P$ that 
\[ r,\pi, \tk,\ttt\models 
\pperm(\tk,\tk_{\tq,\tuW},\tuA) \land (\tuD\rimp \pdel(\tk,\tk_{\tq,\tuW},\tuA)). \]  

This completes the proof that $\pi$ is consistent with $r$. 

To show that $\pi$ is the minimal interpretation consistent with $r$,
suppose that $\pi' = (L', P')$.  
Suppose that $\tk'\in L(\tk,\tn,\ttt)$
and let $\tuW= \tuWt$.  
Then there exist
principal expressions $\tp$ and $\tq$ such that $\tk=\tk_{\tp,\tuWt}$,
$\tk'=\tk_{\tq,\tuW}$ and $\Tuples{C,C_R} \derivp \la \tp\s
\tn,\tq,\tuV' \ra$
for some interval $\tuV'\supseteq \tuW$.  Thus, $r(0)$ contains the
certificate ${\tt (cert~\tk_{\tp,\tuW}~\tn~\tk_{\tq,\tuW}
~\tuW
)}$. This
implies
that if $\pi'=(L',P')$ is another interpretation consistent with $r$
then we must also have $\tk'\in L'(\tk,\tn,\ttt)$.  A similar argument
works in the case of $P$.   Thus, $\pi$ is the minimal interpretation
consistent with $r$.

It remains to show that $r, \tk, \ttt_0 \not \models_c \phi_{\tc}$ for
some choice of $\ttt_0$.  If $\tc$ is a point-valued certificate such
that $\tau_{\tc} = \la \tk,\tn, \tp, [\ttt_0,\ttt_0] \ra$,  
then since it is not the case that 
$\Tuples{C,C_R} \derivq \tau_{\tc}$, 
it is also not the case that 
$\Tuples{C,C_R} \derivp \la \tk\s \tn, \tp,V\ra$ for any interval $V$ containing $\ttt_0$.
(Otherwise, from
Proposition~\ref{prop:three-four},
we would be able to use R0 
and R4(a) 
to derive $\tau_{\tc}$.)
Hence, by Lemma~\ref{3redp}, it follows that
$\tk_{\tp,
\tuW(\ttt_0)} 
\notin L(\tk,\tn,\ttt_0)$.  
Of course, we do have (using R0) that 
$\tk_{\tp,\tuW(\ttt_0)} \in \intension{\tp}{L,\tk, \ttt_0}$.
Thus,  $r,\tk, \ttt_0 \not \models_c \phi_{\tc}$.

Next, suppose that $\tc$ is an arbitrary naming certificate, with
$\tau_{\tc} = \la \tk,\tn,\tp,\tuV \ra$.  Let $\V'$ consist of all
intervals
$\tuV' \in \V$ such that $\Tuples{C,C_R} \derivp \la \tk\s\tn, \tp,
\tuV' \ra$.  $\V'$ is finite, since $\V$ is.  Moreover, it cannot be the
case that $\tuV \subseteq \union \V'$, for otherwise, using R4(a) and
R0,
it would follow that $\Tuples{C,C_R} \derivq \tau_{\tc}$.  Choose
$\ttt_0 \in \tuV - \union \V'$.  It follows just as above that
$r,\tk, \ttt_0 \not \models_c \phi_{\tc}$.

Finally, suppose that $\tc$ is an  authorization certificate, with
$\tau_{\tc} = \la \tk, \tp, \true, \tuA, \tuV\ra$.   (The argument is similar
if $\true$ is replaced by $\false$.)  
We claim that there must be some
time $\ttt_0$ and action $\act \in \aint(\tuA)$ such that for no interval
$\tuV'$ containing $\ttt_0$ and $\tuA'$ 
such that $\act \in \aint(\tuA')$ is it 
the case that $\Tuples{C,C_R} \derivp \la \tk, \tp, \true, \tuA', \tuV'
\ra$.  
It follows from the claim that 
$P(\tk,\ttt_0,\tk_{\tp,\tuW_{\ttt_0}},\act) \ne 2$, so 
$r,\tk, \ttt_0 \not \models_c \phi_{\tc}$.  This completes the proof of
the theorem. 
For the proof of the claim, 
suppose, by way of contradiction, that for all $\ttt\in \tuV$ and all
actions $\act\in \aint(\tuA)$, 
there exists an interval $\tuV_{\ttt,\act}$ containing $\ttt$  and an action
expression $\tuA_{\ttt,\act}$ 
such that  $\act \in \aint(\tuA_{\ttt,\act})$ such that 
$\Tuples{C,C_R} \derivp \la \tk, \tp, \true, \tuA_{\ttt,\act},
\tuV_{\ttt,\act} \ra$. 
Note that there are only finitely many time intervals $\tuV'$ and action
expressions $\tuA'$ such that
$\Tuples{C,C_R} \derivp \la \tk, \tp, \true, \tuA', \tuV' \ra$.
For each $\act \in \aint(\tuA)$, let $\tuA_{\act}$ be the intersection
of all the action expressions $\tuA_{\act,\ttt}$ for $\ttt \in \tuV$.
Since this is a finite intersection, $\tuA_{\act} \in \Ae$.  
Moreover, even if $\aint(\tuA)$ is infinite, the number of distinct sets
$\tuA_{\act}$ is finite.
By R4(c), for each $\ttt \in \tuV$, we have that $\Tuples{C,C_R} \derivq
\la \tk, \tp, \true, \tuA_{\act}, \tuV_{\ttt,\act} \ra$. 
Since the union of the sets $\tuV_{\ttt,\act}$ contains $\tuV$, and
there are only finitely many such sets, by R4(b) it follows that
$\Tuples{C,C_R} \derivq \la \tk, \tp, \true, \tuA_{\act}, \tuV \ra$. 
Finally, since $\aint(\tuA)$ is contained in the union of the sets
$\aint(\tuA_{\act})$, it follows from R4(c) that
$\Tuples{C,C_R} \derivq \la \tk, \tp, \true, \tuA, \tuV \ra$,
which is a contradiction. \eprf
\commentout{
Observe that the union of the set of 
minimal
intervals of this 
sort contains $\tuV$. Let $\tuV'$ be a minimal interval of this sort
and pick a time $\ttt\in \tuV'$. Let $\la \tk, \tp, \true, \tuA_i, \tuV_i \ra$
be the set of tuples promised by the assumption, for the fixed time $\ttt$
as $\act$ ranges over $\tuA$. By the minimality of $\tuV'$, 
for each $i$ we have $\aint(\tuA_i)\supseteq \aint(\tuA')$, so by R4(b)
we obtain that  
$\Tuples{C,C_R} \derivq \la \tk, \tp, \true, \tuA_i, \tuV' \ra$. 
Note that $\bigcup_i \tuA_i \supseteq \tuA$. Hence by R4(c) 
we obtain that $\Tuples{C,C_R} \derivq \la \tk, \tp, \true, \tuA, \tuV' \ra$. 
Using
the observation above about the minimal intervals $\tuV'$
and R4(c), it follows that $\Tuples{C,C_R} \derivq \la \tk, \tp, \true,
\tuA, \tuV \ra$,  
which is a contradiction. 
\eprf 
}
\paragraph{Acknowledgments:}  We thank Jon Howell 
and the reviewers of this paper for their useful comments.
\bibliographystyle{chicago}
\bibliography{joe,z}

\end{document}